\documentclass{aa}  

\usepackage{graphicx}
\usepackage{color}
\usepackage{txfonts}
\newcommand*\rad{~rad\,m$^{-2}$}
\newcommand*\red{\textcolor{black}}

\begin{document}

   \title{Magnetic field at a jet base: extreme Faraday rotation in
     3C\,273 revealed by ALMA }

   \author{T.~Hovatta
     \inst{1,2,3}
     \and
     S. O'Sullivan \inst{4}
     \and
     I. Mart\'i-Vidal \inst{5}
     \and
     T. Savolainen \inst{2,6,7}
     \and
     A. Tchekhovskoy \inst{8,9}
   }

   \institute{Tuorla Observatory, University of Turku,
     V\"ais\"al\"antie 20, FI-21500 Piikki\"o, Finland\\
     \email{talvikki.hovatta@utu.fi}
     \and
     \institute{2}Aalto University Mets\"ahovi Radio Observatory,
     Mets\"ahovintie 114, FI-02540 Kylm\"al\"a, Finland
     \and
     \institute{3}Finnish Centre for Astronomy with ESO, University of
     Turku, FI-20014 University of Turku, Finland
     \and
     \institute{4}Hamburger Sternwarte, Universit\"at Hamburg,
     Gojenbergsweg 112, D-21029 Hamburg, Germany
     \and
     \institute{5}Department of Earth and Space Sciences, Chalmers
     University of Technology, Onsala Space Observatory, SE-439 92,
     Onsala, Sweden
     \and
     \institute{6}Aalto University Department of Electronics and Nanoengineering, PL~15500, FI-00076 Aalto, Finland
     \and
     \institute{7}Max-Planck-Institut f\"ur Radioastronomie, Auf dem H\"ugel 69, D-53121 Bonn, Germany
     \and
     \institute{8}Center for Interdisciplinary Exploration \& Research in Astrophysics (CIERA),
Physics \& Astronomy, Northwestern University, Evanston, IL 60202, USA
     \and 
     \institute{9}
Departments of Astronomy and Physics, Theoretical Astrophysics Center, University of California Berkeley, Berkeley, CA 94720-3411, USA
   }
   
   \date{Received XXX accepted XXX}

 
  \abstract
   {}
   {We  studied the polarization behavior of the quasar 3C\,273 over
     the 1\,mm wavelength band at ALMA with a total bandwidth of
     7.5\,GHz across 223 to 243\,GHz at 0.8\arcsec resolution, corresponding to 2.1\,kpc at the distance of 3C\,273. With these
     observations we were able to probe the optically thin polarized
     emission close to the jet base, and constrain the
     magnetic field structure.}
   {We computed the Faraday rotation measure using
   simple linear fitting and Faraday rotation measure synthesis. In addition, we modeled the broadband behavior of the fractional Stokes Q and U parameters (qu-fitting). The
     systematic uncertainties in the polarization observations at ALMA were
     assessed through Monte Carlo simulations.}
   {We find the unresolved core of 3C\,273 to be \red{1.8\%} linearly
     polarized. We detect a very high rotation measure (RM) of \red{$(5.0\pm0.3)\times10^5$\rad}
     over the 1\,mm band \red{when assuming a single polarized
       component and an external RM screen.} This
     results in a rotation of \red{$>40\degr$} of the intrinsic electric
     vector position angle, which
     is significantly higher than typically assumed for
     millimeter wavelengths. \red{ The polarization fraction increases 
     as a function of wavelength, which according to our 
     qu-fitting could be due to multiple
       polarized components of different Faraday depth within our beam, or internal Faraday
       rotation. With our limited wavelength coverage, we cannot
       distinguish between the cases, and}
     additional multifrequency and high angular resolution observations are
     needed to determine the location and structure of the magnetic field 
     of the Faraday active region. 
     Comparing our RM estimate with values
     obtained at lower frequencies, the RM increases as a
     function of observing frequency, following a power law with an
     index of \red{$2.0\pm0.2$}, consistent with a sheath surrounding a conically expanding jet. We also detect
      $\sim0.2$\% circular polarization, although further
     observations are needed to confirm this result. }
   {}


   \maketitle
%

\section{Introduction}\label{sect:intro}
Magnetic fields are thought to play a significant role in the
formation of relativistic jets in active galactic nuclei
(AGN). Magnetic fields can extract energy from the spinning
supermassive black hole via the Blandford--Znajek mechanism
\citep{blandford77}, launching a magnetically dominated outflow.
In recent years, a major leap forward has been achieved through
general relativistic magnetohydrodynamic (GRMHD) simulations of black
hole accretion that show
how this process can efficiently launch jets \citep[e.g.,][]{2004ApJ...611..977M,2005ApJ...620..878D,2006ApJ...641..103H,mckinney09,tchekhovskoy11}. 

Polarization observations, especially observations of Faraday
rotation, can be used to probe the magnetic fields and magnetized
plasma around relativistic jets, providing the means to connect
observations to the theory of jet formation. When synchrotron emission passes through magnetized plasma, it
undergoes Faraday rotation due to an induced phase offset between the
velocities of the orthogonal plasma modes \citep{burn66}. The amount
of Faraday rotation, the rotation measure (RM), is proportional to the electron density $n_e$ of the plasma (in cm$^{-3}$) and the magnetic
field component $\mathbf{B}$ (in $\mu$G) along the line of sight (in
parsecs), given by
\begin{equation}\label{eq:RM}
\mathrm{RM} = 0.81 \int n_e \mathbf{B} \cdot \mathbf{\mathrm{d}l} ~\mathrm{rad\,m}^{-2}.
\end{equation}

In  the simplest case, Faraday rotation results in a linear dependence
between the observed electric vector position angle (EVPA,
$\chi_\mathrm{obs}$) and wavelength squared ($\lambda^2$) so that
$\chi_\mathrm{obs} = \chi_0 + \mathrm{RM}\lambda^2$, where $\chi_0$ is
the intrinsic electric vector position angle of the emission
region. This allows us to estimate the amount of Faraday rotation by
observing the EVPA at multiple frequencies.

Centimeter-band Faraday rotation observations give us information
about the magnetic field structure around the jets on parsec
scales. The RM of the cores of AGN at centimeter wavelengths are
typically up to a few thousand\rad
\citep[e.g.,][]{taylor98,hovatta12}, indicating magnetic field
strengths of a few $\mu$G if the Faraday rotating screen is the
narrow line region of the source \citep{zavala04}.
In order to study the plasma close to the black hole, millimeter-band
observations are required to probe the jet base in the optically thin
regime \citep[e.g.,][]{lobanov98}. 
\cite{plambeck14} detected a high rotation measure of
$9\times10^5$~\rad\ in  the radio galaxy 3C~84 using observations
from SMA and CARMA. In the low-luminosity AGN M87, a high rotation
measure of RM $< 2 \times 10^5$ ~\rad\ was detected in the SMA
observations by \cite{kuo14}. These observations were interpreted by
assuming that the Faraday rotation occurs in the radiatively
inefficient accretion flow, and both studies estimated the mass accretion rate onto the black
hole. Recently, \cite{moscibrodzka17} showed that in the case of M87,
the polarized emission and Faraday rotation is most likely dominated
by the forward jet instead of the accretion flow, which limits the use
of the RM to probe the accretion rate even in the low-luminosity sources. 

The highest RM detected so far, $2\times10^7$~\rad\ ($\sim 10^8$
~\rad\ in the source frame), was seen in the
high-power gravitationally lensed quasar PKS~1830$-$211 in dual-polarization ALMA
observations by \cite{marti-vidal15}. Based on the frequency-dependence of the rotation measure, they
argued that the Faraday rotation must originate in the jet from a region located 0.01 parsec  from the black
hole. Their result implies that the electron density and/or
magnetic field is high, on the order of  tens of
  Gauss at least, in the jet launching region. 

In this paper we present the first full polarization ALMA observations of the nearest
high-power quasar 3C\,273 ($z = 0.158$; \citealt{strauss92}) over the 1~mm wavelength range. We selected 3C\,273 for
our pilot study because earlier observations of it in the 7mm to 3mm
range have shown indications of $|$RM$| > 2\times10^4$~\rad\
\citep[][Savolainen in prep.]{attridge05,hada16}, making it a good candidate for
high-RM studies in the 1~mm band.

Our paper is organized as follows. The observations and data reduction
are described in Sect.~\ref{sect:obs}. Section~\ref{sect:results}
contains our Faraday rotation and polarization modeling results. We present our discussion
in Sect.~\ref{sect:discussion}, and list our conclusions in Sect.~\ref{sect:conclusions}. We define the
spectral index $\alpha$ such that the total intensity $I$ at frequency
$\nu$ follows the relation $I_\nu \propto \nu^\alpha$.


\section{ALMA observations and data reduction}\label{sect:obs}
The Cycle 4 ALMA observations at band 6 (1.3\,mm) were taken on December 20, 2016, using 44 of the
12m antennas. The observations lasted from 09:32 UTC until 12:05 UTC
with a total integration time of about 73~min on 3C\,273\footnote{A
  second observation to complete the 3-hour program was conducted on December 28, 2016,
  but the duration of the track was only 72 minutes with insufficient
  parallactic angle coverage to calibrate the polarization, which is
  why these data are not used in the analysis.}.
Our observations were taken in full polarization mode using the
recommended continuum setup for band 6, where the four 1.875\,GHz spectral
windows (spw), each consisting of 64 channels that are 31.25\,MHz wide, are
placed at 224, 226, 240, and 242\,GHz. The total bandwidth in our
observations was then 7.5\,GHz. The angular resolution (size of the
synthesized beam) was 0.8\arcsec. The quasar
3C~279 was observed as a bandpass and polarization calibrator and J1224+0330 as the
gain calibrator.

The data reduction was done using Common Astronomical Software
Applications (CASA) version 4.7.0. The amplitude scale of the
observations was set using 3C279, assuming a flux density of 5.86\,Jy
at 233\,GHz with a spectral index of $-0.63$. Calibration of the cross-hand
delay, cross-hand phase, and instrumental polarization (D-terms) was
done using 3C279. For details of the standard data reduction and polarization
calibration steps, see \cite{nagai16}.

\red{From these data with the standard calibration applied, we detect an inconsistency between spw~3
and the rest of the spectral windows, with a discontinuity in
fractional polarization and EVPA that is very likely of instrumental
origin. Since this window is the most affected by atmospheric opacity
(see, e.g. the System temperature plots in Fig.~\ref{Tsys}), so the amplitudes
and phases may be affected by differential atmospheric transmission
between the phase calibrator and the target, we have performed an
additional calibration step to minimize atmospheric and instrumental
biases at the highest frequencies of our observations.}

\red{First, we have self-calibrated the visibilities of our target using the
CASA task \texttt{gaincal} in the so-called ``T'' mode, which ensures
that the same corrections are found and applied to both polarization
channels. Afterwards, we run a bandpass calibration using the target's
visibilities, averaging also the solutions to ``T'' mode, so that the
same correction is applied to both polarizations. After this extra
calibration, we find an increase in the signal-to-noise ratio of the visibilities, as
well as a more continuous behavior of fractional polarization and EVPA
across all spectral windows.}

In our analysis, we assume that the emission in 3C\,273 is
dominated by the central point source at the phase center of the
image. In order to obtain Stokes parameters of each 31.25\,MHz wide
spectral channel, we use the external CASA library {\sc UVMULTIFIT}
\citep{marti-vidal14} to fit the visibility data. We fit a centered
delta component to each channel individually to obtain the Stokes I,
Q, U, and V values. These are listed in Table 1
along with the derived polarization fraction and EVPA.

Following \cite{vlemmings17}, we perform Monte
Carlo (MC) simulations to assess the uncertainties in the Stokes
parameters. These simulations, accounting especially for the
  systematic calibration uncertainties, are described in detail in
Appendix A. We find that the uncertainty is on average \red{8.4\,mJy} in
Stokes I, \red{8.2\,mJy} in Stokes Q, and \red{8.2\,mJy} in Stokes U. This results
in a high accuracy in the fractional polarization \red{($\sim 0.18\%)$},
while the EVPA uncertainty is on the order of 0.5$\degr$. We report
the standard deviation of the 100 MC simulations as the
uncertainty for each parameter in Table 1.

\section{Results}\label{sect:results}
In  this section we describe the methods used to analyze the polarization
data over the ALMA 1\,mm band. Thanks to the wide ALMA bandwidth of
7.5\,GHz spread over a wide frequency range from 223\,GHz to 243\,GHz,
we were able to study the polarization over the single observing band.
We first determined the RM using
multiple methods and then modeled the Stokes I, Q, and U
behavior as a function of wavelength squared using the broadband
polarization modeling technique
known as qu-fitting \citep[e.g.,][]{farnsworth11,osullivan12,farnes14,anderson16,osullivan17,schnitzeler17}. 
In addition to the RM, this method can provide an estimate of the
depolarization in the Faraday medium.

\begin{figure}
   \centering
  \includegraphics[width=\hsize]{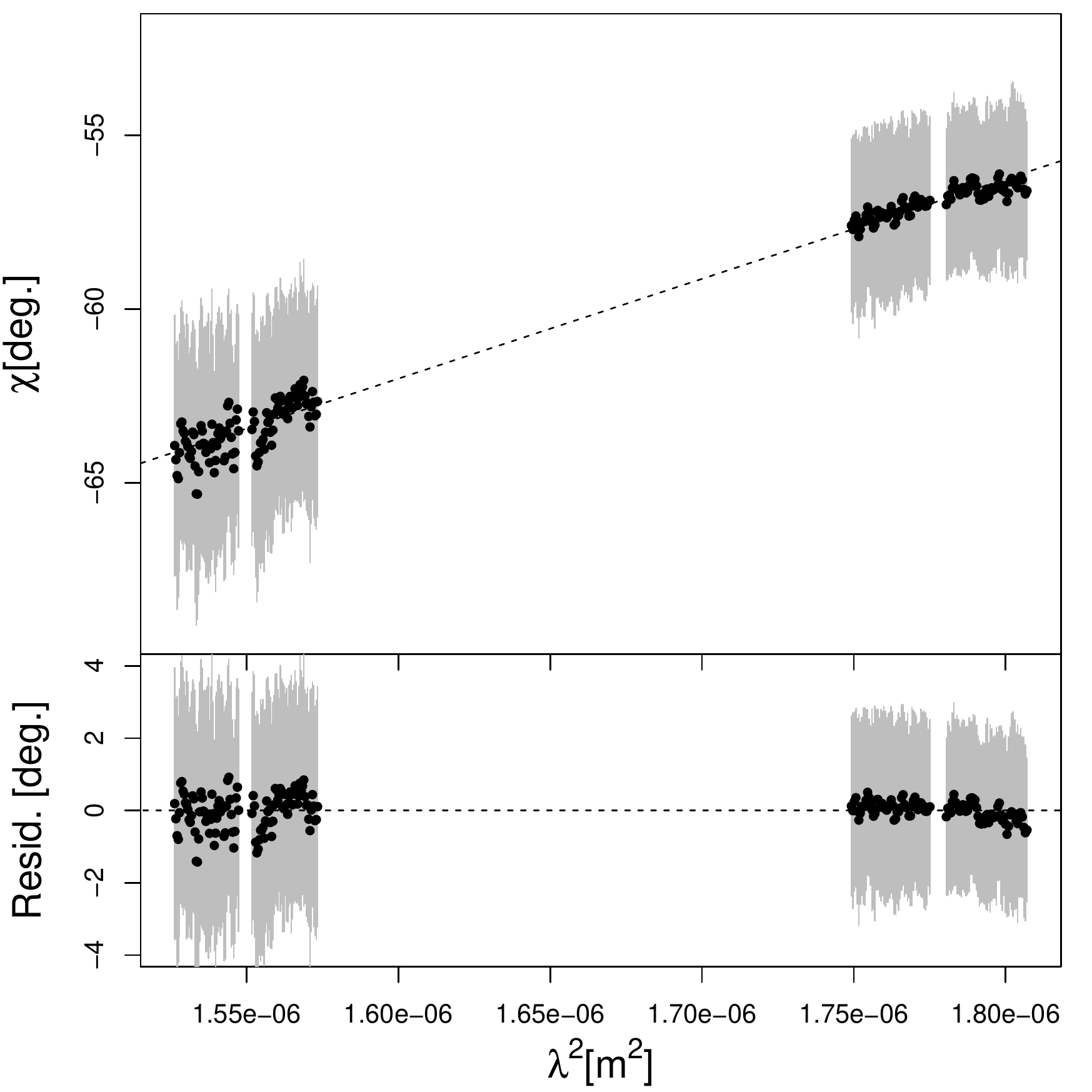}
      \caption{Top: EVPAs as a function of wavelength squared for
        3C\,273 over the ALMA 1\,mm band. The uncertainties in the EVPA are obtained through MC
        simulations described in Appendix A. The dashed line is a
        linear fit to the data giving
        \red{RM$=(5.01\pm0.04)\times10^5$~\rad} and intrinsic EVPA of
        \red{-108$^\circ$}. Bottom: Residuals of the fit with MC uncertainties
        shown in gray.
              }
         \label{fig:RMfit}
   \end{figure}

\subsection{RM determination}
The simplest way to estimate the RM is to directly fit the
EVPAs as a function of wavelength squared. The underlying assumption is that the
RM is caused by a single external Faraday screen, which results in a
linear relation between the EVPAs and wavelength squared. This fit is
shown in Fig.~\ref{fig:RMfit}, and from the slope we obtain 
an \red{RM$=(+5.01\pm0.04)\times10^5$\rad}. The intrinsic EVPA given by
the intercept is \red{-108$^\circ$}, meaning that the Faraday
rotation has a significant effect on the observed EVPAs, contrary to what is  typically assumed at mm wavelengths.

 \begin{figure*}
   \centering
  \includegraphics[width=\hsize]{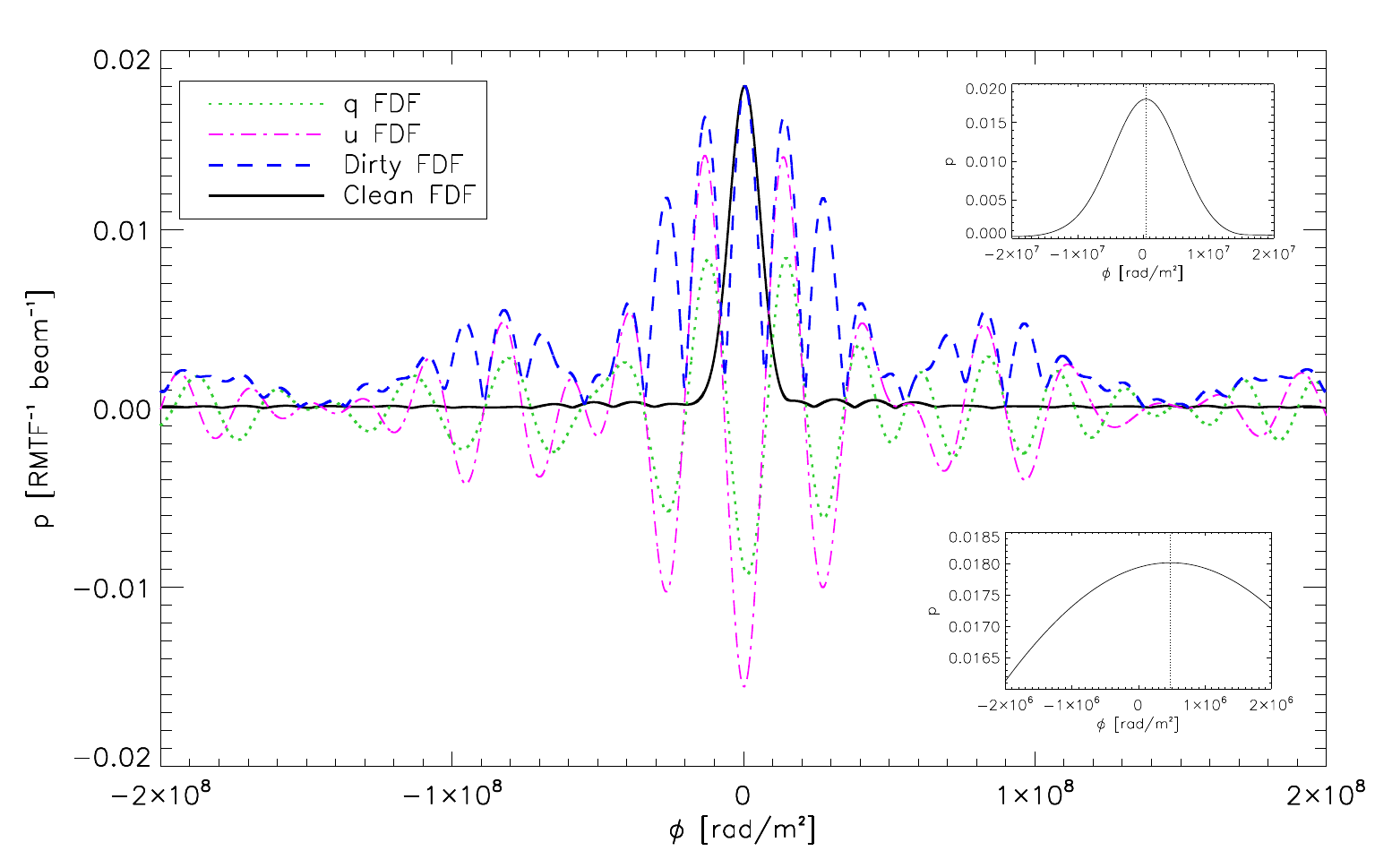}
      \caption{Amplitude of the Faraday dispersion function (FDF) after applying RM synthesis. The Faraday depth ($\phi$) varies 
      from $-2\times10^8$ to $+2\times10^8$\rad~sampled at intervals of $10^4$\rad. The dirty FDFs are shown for $q=Q/I$ 
      (green dotted line), $u=U/I$ (magenta dot-dashed line), and $p$ (blue dashed line);  the FDF after applying RM CLEAN is  also
      shown (solid black line). The inset in the top right is a zoom-in of the main lobe of the FDF, while the bottom right is 
      further zoomed-in to highlight the non-zero Faraday depth of the peak polarized emission (vertical dotted line in both cases). 
              }
         \label{fig:RMsynth}
   \end{figure*}

While the advantage of this method is its
simplicity, we need to manually adjust for any possible $n\pi$
ambiguities in the EVPAs. Looking at the residuals in
Fig.~\ref{fig:RMfit}, we can see that all the spectral windows do
not fall exactly on this line. This may indicate
that the EVPAs do not follow a simple $\lambda^2$ law and 
highlights that the simple method does not give correct results if
there are multiple polarized components that undergo different amounts
of Faraday rotation. These problems can potentially be overcome by using Faraday
rotation measure synthesis \citep{brentjens05}, where we coherently search for 
polarized emission as a function of Faraday depth.

The nominal RM resolution of our observation is $\sim1.3\times10^7$\rad, estimated from the FWHM of the rotation measure spread function (RMSF).\footnote{In practice, an RM lower than this resolution can be identified at high signal-to-noise ratios, analogous to determining the centroid of a radio source at higher precision than the nominal angular resolution.} The maximum RM that can be measured from our data is $\sim\pm\,4\times10^9$\rad~and the maximum scale\footnote{Analogous to the largest angular scale of emission detectable by an interferometer.}  in RM is $\sim2\times10^6$\rad~(meaning that RM structures broader than this will be heavily depolarized). 
By applying RM synthesis 
to the full dataset, we find an RM of
\red{$(+4.8\pm0.2)\times10^5$\rad~}with a degree of polarization of \red{$(1.81\pm0.01)$\% ($\sim$84~mJy~beam$^{-1}$ at 233 GHz)}. We also apply RM CLEAN \citep{heald09} to deconvolve the RM spectrum, but find no additional peaks in the RM spectrum significantly above the band-averaged noise level of $\sim$0.3~mJy~beam$^{-1}$ (Figure~\ref{fig:RMsynth}).

\subsection{Polarization modeling}\label{sect:model}
As described in Sect.~\ref{sect:intro}, in the simplest case, the
Faraday rotation is due to an external screen that does not cause
depolarization. If the Faraday rotating medium is mixed with the
emitting region, or if there are variations in the RM screen over the
finite resolution of the observations, depolarization towards longer
wavelengths may be seen \citep{burn66}. \red{If the Faraday rotating
  medium is mixed with the emitting region (internal Faraday
  rotation), it is also possible to obtain inverse depolarization
  where the polarization increases as a function of wavelength
  \citep{sokoloff98,homan12}. Inverse depolarization can also result
  from multiple emitting RM components \citep{osullivan12}.}

In order to investigate the Faraday depolarization properties of the
data in a quantitative manner, we used the qu-fitting technique. 
\red{We first considered the simplest model, i.e. a model of a
  polarized component in the presence of Faraday rotation, as shown in
  Eq.~\ref{eq:simplemodel} \citep{sokoloff98,osullivan12},}

\begin{figure*}
   \centering
  \includegraphics[width=\hsize]{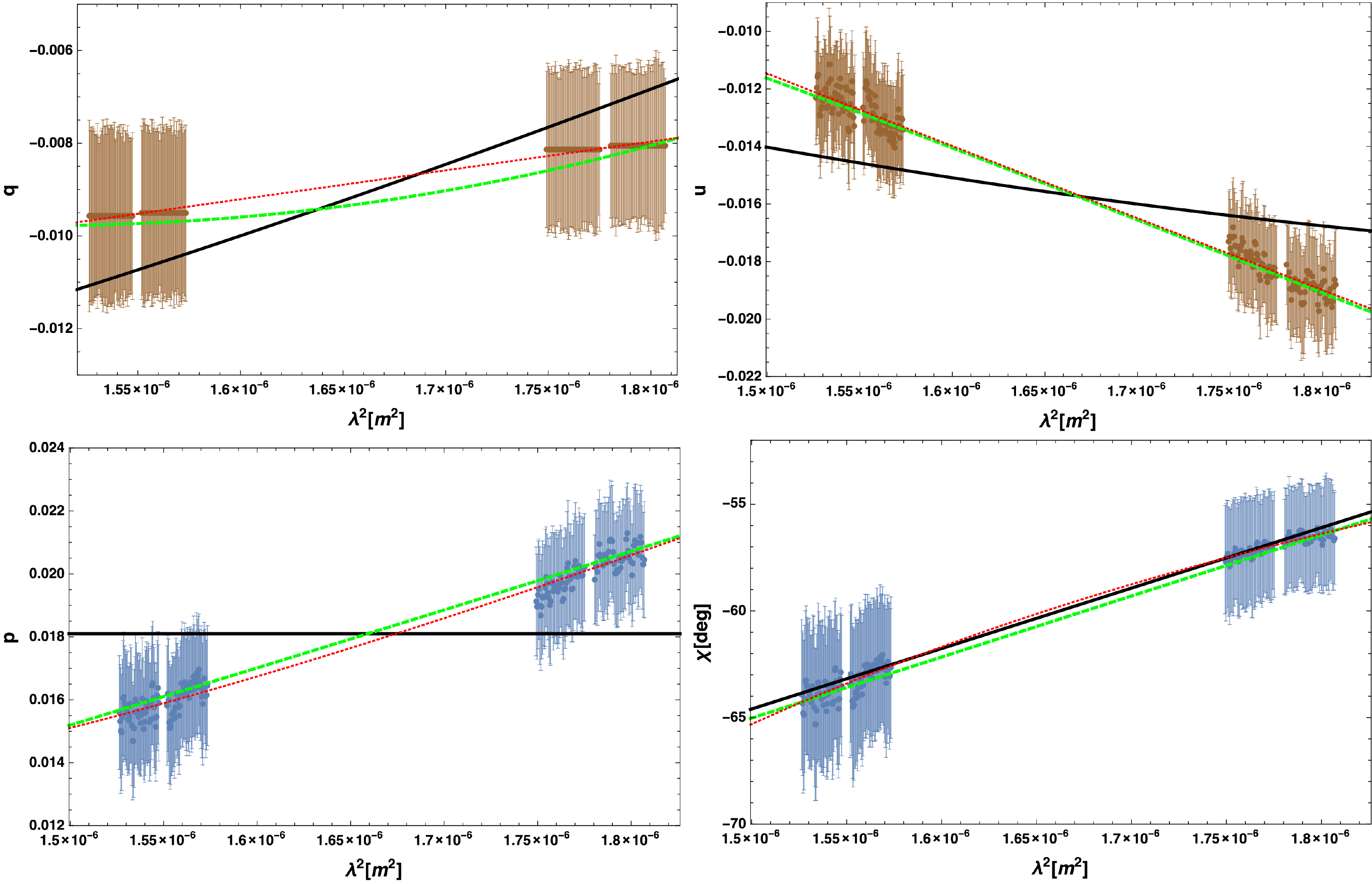}
      \caption{Example fits from the qu-fitting overlaid on the
        data as a function of wavelength squared. Top left: Stokes
        Q/I.  Top right: Stokes U/I. Bottom Left: Fractional
        polarization. Bottom right: EVPA. \red{In
        all panels Model 1 (no depolarization) is shown by a black
        solid line, Model 2 (2 RM components) is shown by a red dotted
        line, and Model 3 (internal Faraday rotation) is shown by a
        green dashed line.}
              }
         \label{fig:qufit}
   \end{figure*}

\red{
\begin{equation}\label{eq:simplemodel}
P=q+iu=p_{0}\,e^{2i(\chi_{0}+{\rm RM}\lambda^2)},
\end{equation}}
\red{where $q=Q/I$, $u=U/I$,  $p_{0}$ is the intrinsic polarization
fraction, $\chi_{0}$ is the intrinsic polarization
angle, and $\lambda$ is the wavelength. This results in
RM=$4.95\pm0.25\times10^5$\rad with a constant $p_{0}=1.81\pm0.01$\% and
$\chi_{0}=-107.1\pm2.3\deg$, consistent with the values obtained from
the simple linear fit shown in Fig.~\ref{fig:RMfit}. This fit is shown as a
black solid line overlaid on the $q$, $u$, $p$, and $\chi$ values in
Fig.~\ref{fig:qufit}. }

\red{However, the behavior of the polarization fraction we observe, where the
  polarization increases with wavelength is not consistent with the
  constant polarization of the simple model. Neither can any single
  component models with external Faraday screens explain such behavior
  \citep{sokoloff98,osullivan12}. Instead, a more complex model with either
  multiple emission components or internal Faraday rotation is
  needed to explain the polarization behavior.}

\red{In case of multiple emission components, the model can simply be
  constructed as $P=P_1 + P_2 + ... + P_N$ \citep{osullivan12}, where
  $P_i$ is as defined in Eq.~\ref{eq:simplemodel}. This kind of a
  2-component  model is shown in Fig.~\ref{fig:qufit} as red dotted lines. The fit shown
  here gives the parameters $p_{01}\sim5.5$\%, $p_{02}\sim4.2$\%,
  $RM1\sim+1.26\times10^5$\rad, $RM2\sim-1.45x10^5$\rad,
  $\chi_{01}\sim88.6\deg$, $\chi_{02}\sim16.7\deg$ for the two
  components. Noteworthy is the smaller absolute RM value and different sign of the
  two components. However, it is possible that also other combinations
  of the parameters, or more than 2 components give equally good fits.}

\red{For the internal Faraday rotation model, we use
 the following equation \citep{sokoloff98,osullivan12}, }
\red{
\begin{equation}\label{eq:internal}
P=p_0\frac{\sin(\Psi-\mathrm{RM_{internal}}\lambda^2)}{\Psi-\mathrm{RM_{internal}}\lambda^2}\,e^{2i(\chi_0+(0.5\mathrm{RM_{internal}}+\mathrm{RM_{external}})\lambda^2)},
\end{equation}}

\red{where $p_0$ and $\chi_0$ are as in Eq.~\ref{eq:simplemodel}, $\Psi$  is the twist of the uniform magnetic field through the jet,
RM$_\mathrm{internal}$ is the amount of internal Faraday rotation, and
RM$_\mathrm{external}$ is the amount of external Faraday rotation. This model, shown in Fig.~\ref{fig:qufit} by green dashed lines, gives us $p_{0}\sim17$\%, $\mathrm{RM_{external}}\sim+3.6\times10^5$\rad,
  RM$_\mathrm{internal}\sim+2.9\times10^5$\rad, $\Psi\sim190\deg$,
  $\chi_0\sim72\deg$. This is again one plausible set of parameters
  that fit the data, but it is not necessarily a unique
  solution. With these parameters, $\mathrm{RM_{external}} + 0.5
  \mathrm{RM_\mathrm{internal}} \sim 5\times10^5$\rad, which is
  consistent with the simple external screen only value.}

\red{However, we emphasize that our current wavelength-squared coverage is not sufficient to uniquely distinguish between these 
two models, or other possible solutions to the models. Moreover, the change in polarization across the band is not very large, and although our MC simulations indicate that it is
significant, given the clear atmospheric signal in spw 3 before the additional calibration steps, these trends should be confirmed with
further data. }

\section{Discussion}\label{sect:discussion}
In the previous section we  show that we detected a Faraday
rotation measure of \red{RM$_\mathrm{obs}\sim+5\times10^5$\rad}~in 3C\,273 over the
1\,mm band ($224-242$\,GHz) of ALMA. Recently, \cite{bower17} reported
1\,mm RM observations of the low-luminosity galaxies M81 and M84
obtained at SMA and CARMA. In their SMA observations, 3C\,273 was used
as a calibrator, and they report a value of RM consistent with zero in
their observations. However, they only used a bandwidth of 4\,GHz in
the RM calculation, which is not wide enough to reveal an RM of \red{$+5
\times 10^5$\rad}~because of their larger uncertainties.

The value we obtain is about an order of magnitude
higher than $|$RM$| > 2\times10^4$~\rad, reported for 3C\,273 in observations between 3\,mm
and 7\,mm \citep[][Savolainen in prep.]{attridge05,hada16},
suggesting a denser Faraday screen or a higher magnetic field
strength over the 1\,mm emission region. \citet{savolainen2008}
estimated the magnetic field strength as a function of distance from
the jet apex in 3C\,273 using multifrequency
Very Long Baseline Interferometry (VLBI) observations. They found the
magnetic field to be $\sim2$\,G in their 3\,mm core, at a distance of $\le0.06$\,mas from the
jet apex. Assuming that the jet is conical and that the magnetic field follows the relation $B\propto r^{-1}$ as
expected for the toroidal component of the field (which could be
expected to correspond to the line-of-sight component of the field),
we would expect the magnetic field strength to be about 6\,G in our observations. Using
Eq.\ref{eq:RM} we can then estimate the electron density in the
Faraday rotating medium. If the path length through the medium is
1\,pc, we obtain \red{$n_e = 10\times10^{-2}$cm$^{-3}$}, which is close to
the central density of $6\times10^{-2}$cm$^{-3}$ found in the X-ray observations of
\cite{roeser00}. 

Alternatively, we can estimate the required magnetic
field strength by assuming a value for the electron
density. A higher electron density of 1000\,cm$^{-3}$, possibly typical of narrow line region clouds \citep{zavala04}, over a
path length of 10\,pc would result in a much lower magnetic field
strength of \red{62\,$\mu$G}. In order to constrain this further, we need to
compare our observation to simulations similar to those of
\cite{moscibrodzka17}, which we plan to do in a forthcoming publication.

\red{ We find that no simple model can explain the polarization
  behavior across the band. Instead, the inverse depolarization could
  be due to two (or more) RM components, or internal Faraday
  rotation. The former case could occur if there are multiple
  strongly polarized components within our beam. The synthesized beam size of our observations is 0.8$\arcsec$, 
which at the distance of 3C\,273 translates to 2.1 kpc, and includes
the entire parsec-scale jet as observed by VLBI. \cite{casadio17}
observed 3C\,273 at 86\,GHz with the Global Millimeter VLBI Array at
50$\mu$as angular resolution in May 2016, about seven months before
our observation. They see only a single dominating polarized component
down the jet. However, earlier 3\,mm VLBI observations have shown
multiple polarized jet components in 3C\,273 \citep[][Savolainen in prep.]{hada16} so that due to possible variability of the source, we
cannot exclude the possibility of multiple polarized components. }

\red{In case of internal Faraday rotation, an increasing
  polarization as a function of wavelength could be obtained, for
  example, if there is a helical magnetic field within the jet
  \citep{homan12}. This kind of inverse depolarization has been
  observed also in the pc-scale jet of 3C\,273 \citep{hovatta12}. 
A helical field is also supported by observations of a parsec-scale
transverse Faraday rotation measure gradient transverse to the jet
direction in 3C\,273
\citep[e.g.,][]{asada02,zavala05,hovatta12}. An ordered, helical
magnetic field would be expected, for example, if the black hole is surrounded by dynamically important poloidal magnetic
fields \citep{tchekhovskoy11} that wind up to a tight helix due to the
rotation of the black hole / accretion disk system \citep[e.g.,][]{meier01}.}

\red{We emphasize that because of the limited wavelength coverage and angular resolution, 
our observations cannot distinguish between the two cases, and further
observations with improved angular resolution such as observations
with the Event Horizon Telescope \citep{doeleman09} and Global
Millimeter VLBI Array (GMVA) + ALMA \citep[e.g.,][]{boccardi17} are
needed to help determine the origin and properties of this
Faraday rotation medium. Alternatively, observations at multiple bands
will constrain the qu-fitting more, providing more strict constraints
for the models.}

\subsection{RM as a function of frequency}\label{sect:rmfreq}
The behavior of RM as a function of frequency can tell us about the
underlying conditions in the jet as both the electron density and
magnetic field strength change as a function of distance from the
black hole, which under certain assumptions can be translated to
observing frequency. As derived in \cite{jorstad07}, in
a conical jet under equipartition the RM is expected to follow a
relation $|$RM$| \propto \nu^a$, where the value of $a$ depends on the
power-law change in the electron density $n_e$ as a function of
distance $r$ from the black hole, $n_e \propto r^{-a}$.
For example, $a = 2$ would imply that the Faraday rotation
is occurring in a sheath around a conically expanding jet, while
lower values could be explained with a more highly collimated jet
\citep[e.g.,][]{osullivan09a}. The typical values for $a$ obtained in
the literature vary in the range
$0.9-4$ with average or median values around 2
\citep{jorstad07,osullivan09a,kravchenko17}. 

In order to look for the frequency dependency in 3C\,273, we gathered
from the literature previously reported values for the core RM at
different frequency ranges. As the core in 3C\,273 at lower frequencies
is typically depolarized, we  selected the value nearest to the
core, so that these should formally be considered as lower limits.  If
multiple values are reported in the literature, we take their mean as
the RM value and use the range as an uncertainty in the calculation of
the slope. We used the range ${\rm RM}=250 - 450$\rad~in the frequency
range $4.7-8$\,GHz from \cite{asada02}. This value is taken about
5\,mas from the core and therefore is probably much lower than
expected for the core.
For the frequency range $8-15$\,GHz we use the range of
near-core RM values of  ${\rm RM}=1000-3000$\rad~reported in \cite{zavala05} and
\cite{hovatta12}. For the $43-86$\,GHz range we use the range of lower
limits $2.1-2.4 \times 10^4$\rad~from \cite{attridge05} and
\cite{hada16}. 

  \begin{figure}
   \centering
  \includegraphics[width=\hsize]{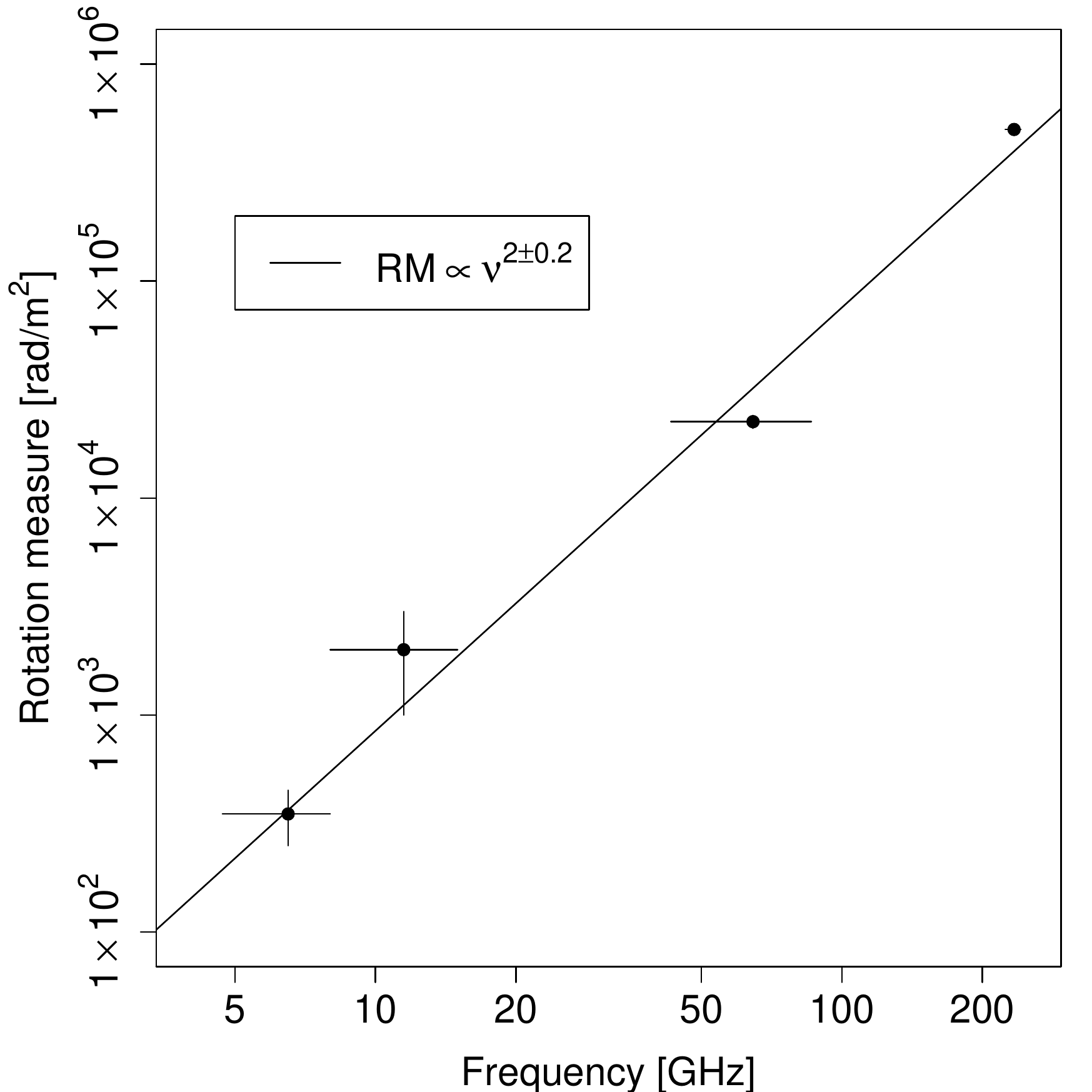}
      \caption{Rotation measure as a function of frequency for
        3C\,273. The RM values at lower than ALMA $224-242$\,GHz
        frequency from this study are taken from the near-core region
        in VLBA observations from the literature (see text for
        references). Apart from the ALMA observation, the uncertainty in
        RM is a range of reported values, not an uncertainty in a
        single measurement. Similarly, the frequency range of the
        observations over the which the RM was estimated is shown by a
        solid line. 
              }
         \label{fig:RMfreq}
   \end{figure}

The linear fit to these data is shown in Fig.~\ref{fig:RMfreq}. We
obtain \red{$a = 2.0\pm0.2$}, consistent with the mean value of $a = 1.8\pm0.2$ by \cite{jorstad07},
and \red{as expected} for a sheath surrounding a
conically expanding flow. However, this slope should be considered as
an upper limit because the RM values at the lower frequencies are most
certainly lower limits. Our result is also consistent with the analysis
presented in \cite{plambeck14}, who found the extrapolation from lower
frequencies with a  slope of $a=2$ to agree with with their high RM of
$9\times10^5$~\rad\ detected at 1.3\,mm in 3C\,84.

At 1\,mm wavelength, the emission in 3C\,273 is
already optically thin \citep[e.g.,][]{courvoisier98, planck11} so
that we may be viewing the polarized emission very close to the black
hole. For example, in the models by \cite{marscher80} and
\cite{potter12}, the emission at optically thin millimeter-band
frequencies originates from the region where the jet transitions from
a parabolic, magnetically dominated jet to a conical jet in
equipartition (referred to as the transition region later in the text).  This is supported by simulations of
\cite{porth11}, who show that at higher, optically thin frequencies the
polarized flux from the unresolved core (jet base in case of optically
thin emission) begins to dominate the RM signal, which
can reach values up to $10^6$\rad, similar to what we
detect. 

However, it is also possible that the polarized
emission is dominated by a single polarized jet component  farther down
the jet as seen in a 86\,GHz VLBI image of 3C\,273 taken in May 2016
\citep{casadio17}. In their image, the core is depolarized, possibly
due to opacity, because at 86\,GHz, 3C\,273 is in the transition phase  to becoming
optically thin. The image of \cite{casadio17} was taken 3 months after a peak of a
large total flux density flare\footnote{See the ALMA Calibrator
  Source Catalogue at \url{https://almascience.eso.org/sc/} for the total flux density evolution of the source.}, and the polarized component may be
related to this flaring activity. The peak polarized flux density detected by
\cite{casadio17} is about 200\,mJy/beam. Assuming a spectral
index of \red{$-0.9$ (as seen over our 1\,mm band), we would expect to
  see about 80\,mJy of polarized flux in our 1\,mm observations, which
  is very close to the mean polarized flux density of 83\,mJy we
  observe. Thus, it is possible that we are seeing a polarized
  component farther down the jet. However, our observations were taken 7 months after the VLBI observations during
a dip in the total flux density curve, and just before a new flare started
to rise so that it is also possible that we are seeing emission
related to the new flaring activity, closer to the transition region,
or multiple emission components, as suggested by our qu-fitting results.}

If at 1\,mm wavelength we are seeing the optically thin emission in the jet
transition region, observations at even higher frequencies should
result in similar RM values to those we have now obtained, assuming that
the Faraday screen is stable. Simultaneous observations at
multiple frequencies will help to answer this question.

 Our observed RM$_\mathrm{obs}$ of \red{$5.0\times10^5$\rad} corresponds to
  \red{RM$_\mathrm{int} = \mathrm{RM}_\mathrm{obs}(1+z)^2=6.7\times10^5$\rad} in the frame of
  the source, at an emitted frequency of
  $\nu_\mathrm{em}=\nu_\mathrm{obs}(1+z)=270$\,GHz when using the
  central observed frequency $\nu_\mathrm{obs}=234$\,GHz. This is
  still much lower than the intrinsic RM of $10^8$\rad\ seen
  in the quasar PKS~1830$-$211 by \cite{marti-vidal15}. This could
  partially be explained with the higher redshift ($z=2.5$) of
  PKS~1830$-$211, which makes the emitted frequency over which the RM
  was detected correspond to 875$-$1050\,GHz, and which could possibly
  originate from closer to the black hole where the electron density
  and magnetic field is expected to be higher. If we estimate the
  expected RM at these high frequencies using the slope \red{$2.0$} from
  above, we obtain an intrinsic RM of \red{$\sim7\times10^6$\rad}, suggesting
  that the reason for the lower RM in 3C\,273 is due to intrinsic
  differences in the electron density and/or magnetic field strength
  in these two sources. Studying a larger number of sources will allow
  us to establish whether this type of difference is common between
  various AGN.

\subsection{Circular polarization}
Although circular polarization observations are not yet officially offered by
the ALMA Observatory, our Monte Carlo uncertainty assessment, described in Appendix A, 
allows us to investigate the Stokes V signal in our data. 
Assuming that we have accounted
for all the possible errors in the data, we find that the instrumental
contribution to the total Stokes V is about $2-3$\,mJy, while there
seems to be an intrinsic signal of about 10\,mJy across the band (see
Appendix A for details). This
would indicate a fractional circular polarization of $\sim
0.2$\%. This result should be treated with caution, and we would need
additional ALMA observations (with properly supported circular
polarimetry) to verify the detection. However, the value we obtain is
consistent with the 1.3\,mm circular polarization observations by the
POLAMI group \citep{thum18} who typically do not detect significant
circular polarization from 3C\,273, which is then expected as their
uncertainties are typically higher than 0.2\%. 

Assuming a magnetic field strength of $\sim$6\,G in the region 
dominating the mm emission and a Doppler factor of $\sim5$ \citep{savolainen2008},
an intrinsic circular polarization as high as 1.8\% can be obtained for a completely 
uniform jet magnetic field. Alternatively, if Faraday rotation-driven conversion 
is dominating the production of circular polarization, then the observed Stokes V level of $\sim$0.2\% 
could be obtained from 5\% linear polarization with a low-energy relativistic electron 
energy spectrum cutoff of $\sim$4.5 \citep[e.g.,][]{homan2009, osullivan2013}. This would only provide 
an internal Faraday depth on the  order of tens of \rad,~and thus indicates that the majority of the 
Faraday rotating material we detect is likely in a boundary layer or wind external to the jet.

\section{Conclusions}\label{sect:conclusions}
We have studied the quasar 3C\,273 at 1\,mm wavelength with ALMA in
full polarization mode at 0.8\,mas resolution, corresponding to
2.1\,kpc at the distance of 3C\,273. We detect about \red{1.8\%} linear polarization in the
unresolved core of the source, and model the polarization as a function
of wavelength over the 1\,mm band. Our main conclusions can be
summarized as follows: 
   \begin{enumerate}
      \item We detect a very high Faraday rotation measure (RM) of
        \red{$(5.0\pm0.3)\times10^5$\rad}~over the band, which  implies a magnetic
        field of several Gauss or a high electron density of
        $\sim1000$~cm$^{-3}$ in the Faraday rotating medium probed 
        by the polarized emission. This amount of Faraday rotation
        rotates the EVPAs in the 1\,mm wavelength by over \red{40$\degr$}, showing
        that it cannot be ignored when studying the intrinsic EVPA and
        magnetic field direction.
      \item By modeling the Stokes parameters as a function of
        wavelength, we find that \red{no model with a single polarized
          component and an external screen can explain the inverse
          depolarization we observe. Instead, a model with at least
          two polarized components, or internal Faraday rotation is
          needed.} Additional multifrequency and high angular 
        resolution observations are
        required to distinguish between the models.
      \item Comparing the RM at 1\,mm to values obtained at lower
        frequencies, we find that the RM  increases as a
     function of observing frequency, following a power law with an
     index of \red{$<2.0\pm0.2$}, consistent with a sheath surrounding a
     conically expanding jet. 
     \item Through careful Monte Carlo assessment of the systematic
       uncertainties in the polarization observations, we are able to 
       detect about 0.2\% of circular polarization. Additional dedicated circular
       polarization observations are needed to confirm the result.
   \end{enumerate}

\begin{acknowledgements}
\red{We thank Prof. Seiji Kameno for providing us with independently
  calibrated polarization measurements for cross-checking our results.}
This paper makes use of the following ALMA data: ADS/JAO.ALMA\#2016.1.01073.S. ALMA 
is a partnership of ESO (representing its member states), NSF (USA), and NINS (Japan), 
together with NRC (Canada), NSC and ASIAA (Taiwan), and KASI (Republic of Korea), in 
cooperation with the Republic of Chile. The Joint ALMA Observatory is operated by 
ESO, AUI/NRAO, and NAOJ. This publication has received funding from the European 
Union’s Horizon 2020 research and innovation programme under grant agreement 
No 730562 [RadioNet]. TH was supported by the Turku Collegium of
Science and Medicine \red{and the Academy of Finland project 317383}. TS was funded by the Academy of Finland projects 274477,
284495, and 312496. AT was supported by the TAC fellowship.
\end{acknowledgements}


\bibliographystyle{aa} 
\bibliography{references.bib} 

\begin{thebibliography}{51}
\expandafter\ifx\csname natexlab\endcsname\relax\def\natexlab#1{#1}\fi

\bibitem[{{Anderson} {et~al.}(2016){Anderson}, {Gaensler}, \&
  {Feain}}]{anderson16}
{Anderson}, C.~S., {Gaensler}, B.~M., \& {Feain}, I.~J. 2016, \apj, 825, 59

\bibitem[{{Asada} {et~al.}(2002){Asada}, {Inoue}, {Uchida}, {Kameno},
  {Fujisawa}, {Iguchi}, \& {Mutoh}}]{asada02}
{Asada}, K., {Inoue}, M., {Uchida}, Y., {et~al.} 2002, \pasj, 54, L39

\bibitem[{{Attridge} {et~al.}(2005){Attridge}, {Wardle}, \&
  {Homan}}]{attridge05}
{Attridge}, J.~M., {Wardle}, J.~F.~C., \& {Homan}, D.~C. 2005, \apjl, 633, L85

\bibitem[{{Blandford} \& {Znajek}(1977)}]{blandford77}
{Blandford}, R.~D. \& {Znajek}, R.~L. 1977, \mnras, 179, 433

\bibitem[{{Boccardi} {et~al.}(2017){Boccardi}, {Krichbaum}, {Ros}, \&
  {Zensus}}]{boccardi17}
{Boccardi}, B., {Krichbaum}, T.~P., {Ros}, E., \& {Zensus}, J.~A. 2017, \aapr,
  25, 4

\bibitem[{{Bower} {et~al.}(2017){Bower}, {Dexter}, {Markoff}, {Rao}, \&
  {Plambeck}}]{bower17}
{Bower}, G.~C., {Dexter}, J., {Markoff}, S., {Rao}, R., \& {Plambeck}, R.~L.
  2017, \apjl, 843, L31

\bibitem[{{Brentjens} \& {de Bruyn}(2005)}]{brentjens05}
{Brentjens}, M.~A. \& {de Bruyn}, A.~G. 2005, \aap, 441, 1217

\bibitem[{{Burn}(1966)}]{burn66}
{Burn}, B.~J. 1966, \mnras, 133, 67

\bibitem[{{Casadio} {et~al.}(2017){Casadio}, {Krichbaum}, {Marscher},
  {Jorstad}, {G{\'o}mez}, {Agudo}, {Bach}, {Kim}, {Hodgson}, \&
  {Zensus}}]{casadio17}
{Casadio}, C., {Krichbaum}, T., {Marscher}, A., {et~al.} 2017, Galaxies, 5, 67

\bibitem[{{Courvoisier}(1998)}]{courvoisier98}
{Courvoisier}, T.~J.-L. 1998, \aapr, 9, 1

\bibitem[{{De Villiers} {et~al.}(2005){De Villiers}, {Hawley}, {Krolik}, \&
  {Hirose}}]{2005ApJ...620..878D}
{De Villiers}, J.-P., {Hawley}, J.~F., {Krolik}, J.~H., \& {Hirose}, S. 2005,
  \apj, 620, 878

\bibitem[{{Doeleman} {et~al.}(2009){Doeleman}, {Agol}, {Backer}, {Baganoff},
  {Bower}, {Broderick}, {Fabian}, {Fish}, {Gammie}, {Ho}, {Honman},
  {Krichbaum}, {Loeb}, {Marrone}, {Reid}, {Rogers}, {Shapiro}, {Strittmatter},
  {Tilanus}, {Weintroub}, {Whitney}, {Wright}, \& {Ziurys}}]{doeleman09}
{Doeleman}, S., {Agol}, E., {Backer}, D., {et~al.} 2009, in , astro2010: The
  Astronomy and Astrophysics Decadal Survey, arXiv:0906.3899

\bibitem[{{Farnes} {et~al.}(2014){Farnes}, {Gaensler}, \&
  {Carretti}}]{farnes14}
{Farnes}, J.~S., {Gaensler}, B.~M., \& {Carretti}, E. 2014, \apjs, 212, 15

\bibitem[{{Farnsworth} {et~al.}(2011){Farnsworth}, {Rudnick}, \&
  {Brown}}]{farnsworth11}
{Farnsworth}, D., {Rudnick}, L., \& {Brown}, S. 2011, \aj, 141, 191

\bibitem[{{Hada} {et~al.}(2016){Hada}, {Kino}, {Doi}, {Nagai}, {Honma},
  {Akiyama}, {Tazaki}, {Lico}, {Giroletti}, {Giovannini}, {Orienti}, \&
  {Hagiwara}}]{hada16}
{Hada}, K., {Kino}, M., {Doi}, A., {et~al.} 2016, \apj, 817, 131

\bibitem[{{Hawley} \& {Krolik}(2006)}]{2006ApJ...641..103H}
{Hawley}, J.~F. \& {Krolik}, J.~H. 2006, \apj, 641, 103

\bibitem[{{Heald} {et~al.}(2009){Heald}, {Braun}, \& {Edmonds}}]{heald09}
{Heald}, G., {Braun}, R., \& {Edmonds}, R. 2009, \aap, 503, 409

\bibitem[{{Homan}(2012)}]{homan12}
{Homan}, D.~C. 2012, \apjl, 747, L24

\bibitem[{{Homan} {et~al.}(2009){Homan}, {Lister}, {Aller}, {Aller}, \&
  {Wardle}}]{homan2009}
{Homan}, D.~C., {Lister}, M.~L., {Aller}, H.~D., {Aller}, M.~F., \& {Wardle},
  J.~F.~C. 2009, ApJ, 696, 328

\bibitem[{{Hovatta} {et~al.}(2012){Hovatta}, {Lister}, {Aller}, {Aller},
  {Homan}, {Kovalev}, {Pushkarev}, \& {Savolainen}}]{hovatta12}
{Hovatta}, T., {Lister}, M.~L., {Aller}, M.~F., {et~al.} 2012, \aj, 144, 105

\bibitem[{{Jorstad} {et~al.}(2007){Jorstad}, {Marscher}, {Stevens}, {Smith},
  {Forster}, {Gear}, {Cawthorne}, {Lister}, {Stirling}, {G{\'o}mez}, {Greaves},
  \& {Robson}}]{jorstad07}
{Jorstad}, S.~G., {Marscher}, A.~P., {Stevens}, J.~A., {et~al.} 2007, \aj, 134,
  799

\bibitem[{{Kravchenko} {et~al.}(2017){Kravchenko}, {Kovalev}, \&
  {Sokolovsky}}]{kravchenko17}
{Kravchenko}, E.~V., {Kovalev}, Y.~Y., \& {Sokolovsky}, K.~V. 2017, \mnras,
  467, 83

\bibitem[{{Kuo} {et~al.}(2014){Kuo}, {Asada}, {Rao}, {Nakamura}, {Algaba},
  {Liu}, {Inoue}, {Koch}, {Ho}, {Matsushita}, {Pu}, {Akiyama}, {Nishioka}, \&
  {Pradel}}]{kuo14}
{Kuo}, C.~Y., {Asada}, K., {Rao}, R., {et~al.} 2014, \apjl, 783, L33

\bibitem[{{Lobanov}(1998)}]{lobanov98}
{Lobanov}, A.~P. 1998, \aaps, 132, 261

\bibitem[{Marscher(1980)}]{marscher80}
Marscher, A.~P. 1980, \apj, 235, 386

\bibitem[{{Mart{\'{\i}}-Vidal} {et~al.}(2015){Mart{\'{\i}}-Vidal}, {Muller},
  {Vlemmings}, {Horellou}, \& {Aalto}}]{marti-vidal15}
{Mart{\'{\i}}-Vidal}, I., {Muller}, S., {Vlemmings}, W., {Horellou}, C., \&
  {Aalto}, S. 2015, Science, 348, 311

\bibitem[{{Mart{\'{\i}}-Vidal} {et~al.}(2014){Mart{\'{\i}}-Vidal}, {Vlemmings},
  {Muller}, \& {Casey}}]{marti-vidal14}
{Mart{\'{\i}}-Vidal}, I., {Vlemmings}, W.~H.~T., {Muller}, S., \& {Casey}, S.
  2014, \aap, 563, A136

\bibitem[{{McKinney} \& {Blandford}(2009)}]{mckinney09}
{McKinney}, J.~C. \& {Blandford}, R.~D. 2009, \mnras, 394, L126

\bibitem[{{McKinney} \& {Gammie}(2004)}]{2004ApJ...611..977M}
{McKinney}, J.~C. \& {Gammie}, C.~F. 2004, \apj, 611, 977

\bibitem[{{Meier} {et~al.}(2001){Meier}, {Koide}, \& {Uchida}}]{meier01}
{Meier}, D.~L., {Koide}, S., \& {Uchida}, Y. 2001, Science, 291, 84

\bibitem[{{Mo{\'s}cibrodzka} {et~al.}(2017){Mo{\'s}cibrodzka}, {Dexter},
  {Davelaar}, \& {Falcke}}]{moscibrodzka17}
{Mo{\'s}cibrodzka}, M., {Dexter}, J., {Davelaar}, J., \& {Falcke}, H. 2017,
  \mnras, 468, 2214

\bibitem[{{Nagai} {et~al.}(2016){Nagai}, {Nakanishi}, {Paladino}, {Hull},
  {Cortes}, {Moellenbrock}, {Fomalont}, {Asada}, \& {Hada}}]{nagai16}
{Nagai}, H., {Nakanishi}, K., {Paladino}, R., {et~al.} 2016, \apj, 824, 132

\bibitem[{{O'Sullivan} {et~al.}(2012){O'Sullivan}, {Brown}, {Robishaw},
  {Schnitzeler}, {McClure-Griffiths}, {Feain}, {Taylor}, {Gaensler},
  {Landecker}, {Harvey-Smith}, \& {Carretti}}]{osullivan12}
{O'Sullivan}, S.~P., {Brown}, S., {Robishaw}, T., {et~al.} 2012, \mnras, 421,
  3300

\bibitem[{{O'Sullivan} \& {Gabuzda}(2009)}]{osullivan09a}
{O'Sullivan}, S.~P. \& {Gabuzda}, D.~C. 2009, \mnras, 393, 429

\bibitem[{{O'Sullivan} {et~al.}(2013){O'Sullivan}, {McClure-Griffiths},
  {Feain}, {Gaensler}, \& {Sault}}]{osullivan2013}
{O'Sullivan}, S.~P., {McClure-Griffiths}, N.~M., {Feain}, I.~J., {Gaensler},
  B.~M., \& {Sault}, R.~J. 2013, \mnras, 435, 311

\bibitem[{{O'Sullivan} {et~al.}(2017){O'Sullivan}, {Purcell}, {Anderson},
  {Farnes}, {Sun}, \& {Gaensler}}]{osullivan17}
{O'Sullivan}, S.~P., {Purcell}, C.~R., {Anderson}, C.~S., {et~al.} 2017,
  \mnras, 469, 4034

\bibitem[{{Plambeck} {et~al.}(2014){Plambeck}, {Bower}, {Rao}, {Marrone},
  {Jorstad}, {Marscher}, {Doeleman}, {Fish}, \& {Johnson}}]{plambeck14}
{Plambeck}, R.~L., {Bower}, G.~C., {Rao}, R., {et~al.} 2014, \apj, 797, 66

\bibitem[{{Planck Collaboration} {et~al.}(2011){Planck Collaboration},
  {Aatrokoski}, {Ade}, {Aghanim}, {Aller}, {Aller}, {Angelakis}, {Arnaud},
  {Ashdown}, {Aumont}, \& et~al.}]{planck11}
{Planck Collaboration}, {Aatrokoski}, J., {Ade}, P.~A.~R., {et~al.} 2011, \aap,
  536, A15

\bibitem[{{Porth} {et~al.}(2011){Porth}, {Fendt}, {Meliani}, \&
  {Vaidya}}]{porth11}
{Porth}, O., {Fendt}, C., {Meliani}, Z., \& {Vaidya}, B. 2011, \apj, 737, 42

\bibitem[{{Potter} \& {Cotter}(2012)}]{potter12}
{Potter}, W.~J. \& {Cotter}, G. 2012, \mnras, 423, 756

\bibitem[{{R{\"o}ser} {et~al.}(2000){R{\"o}ser}, {Meisenheimer}, {Neumann},
  {Conway}, \& {Perley}}]{roeser00}
{R{\"o}ser}, H.-J., {Meisenheimer}, K., {Neumann}, M., {Conway}, R.~G., \&
  {Perley}, R.~A. 2000, \aap, 360, 99

\bibitem[{{Savolainen} {et~al.}(2008){Savolainen}, {Wiik}, {Valtaoja}, \&
  {Tornikoski}}]{savolainen2008}
{Savolainen}, T., {Wiik}, K., {Valtaoja}, E., \& {Tornikoski}, M. 2008, in
  Astronomical Society of the Pacific Conference Series, Vol. 386,
  Extragalactic Jets: Theory and Observation from Radio to Gamma Ray, ed. T.~A.
  {Rector} \& D.~S. {De Young}, 451

\bibitem[{{Schnitzeler}(2018)}]{schnitzeler17}
{Schnitzeler}, D.~H.~F.~M. 2018, \mnras, 474, 300

\bibitem[{{Sokoloff} {et~al.}(1998){Sokoloff}, {Bykov}, {Shukurov},
  {Berkhuijsen}, {Beck}, \& {Poezd}}]{sokoloff98}
{Sokoloff}, D.~D., {Bykov}, A.~A., {Shukurov}, A., {et~al.} 1998, \mnras, 299,
  189

\bibitem[{{Strauss} {et~al.}(1992){Strauss}, {Huchra}, {Davis}, {Yahil},
  {Fisher}, \& {Tonry}}]{strauss92}
{Strauss}, M.~A., {Huchra}, J.~P., {Davis}, M., {et~al.} 1992, \apjs, 83, 29

\bibitem[{{Taylor}(1998)}]{taylor98}
{Taylor}, G.~B. 1998, \apj, 506, 637

\bibitem[{{Tchekhovskoy} {et~al.}(2011){Tchekhovskoy}, {Narayan}, \&
  {McKinney}}]{tchekhovskoy11}
{Tchekhovskoy}, A., {Narayan}, R., \& {McKinney}, J.~C. 2011, \mnras, 418, L79

\bibitem[{{Thum} {et~al.}(2018){Thum}, {Agudo}, {Molina}, {Casadio},
  {G{\'o}mez}, {Morris}, {Ramakrishnan}, \& {Sievers}}]{thum18}
{Thum}, C., {Agudo}, I., {Molina}, S.~N., {et~al.} 2018, \mnras, 473, 2506

\bibitem[{{Vlemmings} {et~al.}(2017){Vlemmings}, {Khouri}, {Marti-Vidal},
  {Tafoya}, {Baudry}, {Etoka}, {Humphreys}, {Jones}, {Kemball}, {O'Gorman},
  {Perez-Sanchez}, \& {Richards}}]{vlemmings17}
{Vlemmings}, W.~H.~T., {Khouri}, T., {Marti-Vidal}, I., {et~al.} 2017, \aap,
  603, A92

\bibitem[{{Zavala} \& {Taylor}(2004)}]{zavala04}
{Zavala}, R.~T. \& {Taylor}, G.~B. 2004, \apj, 612, 749

\bibitem[{{Zavala} \& {Taylor}(2005)}]{zavala05}
{Zavala}, R.~T. \& {Taylor}, G.~B. 2005, \apjl, 626, L73

\end{thebibliography}

\begin{appendix} 
\section{Assessment of the polarization calibration}
\begin{figure}
\includegraphics[width=\hsize]{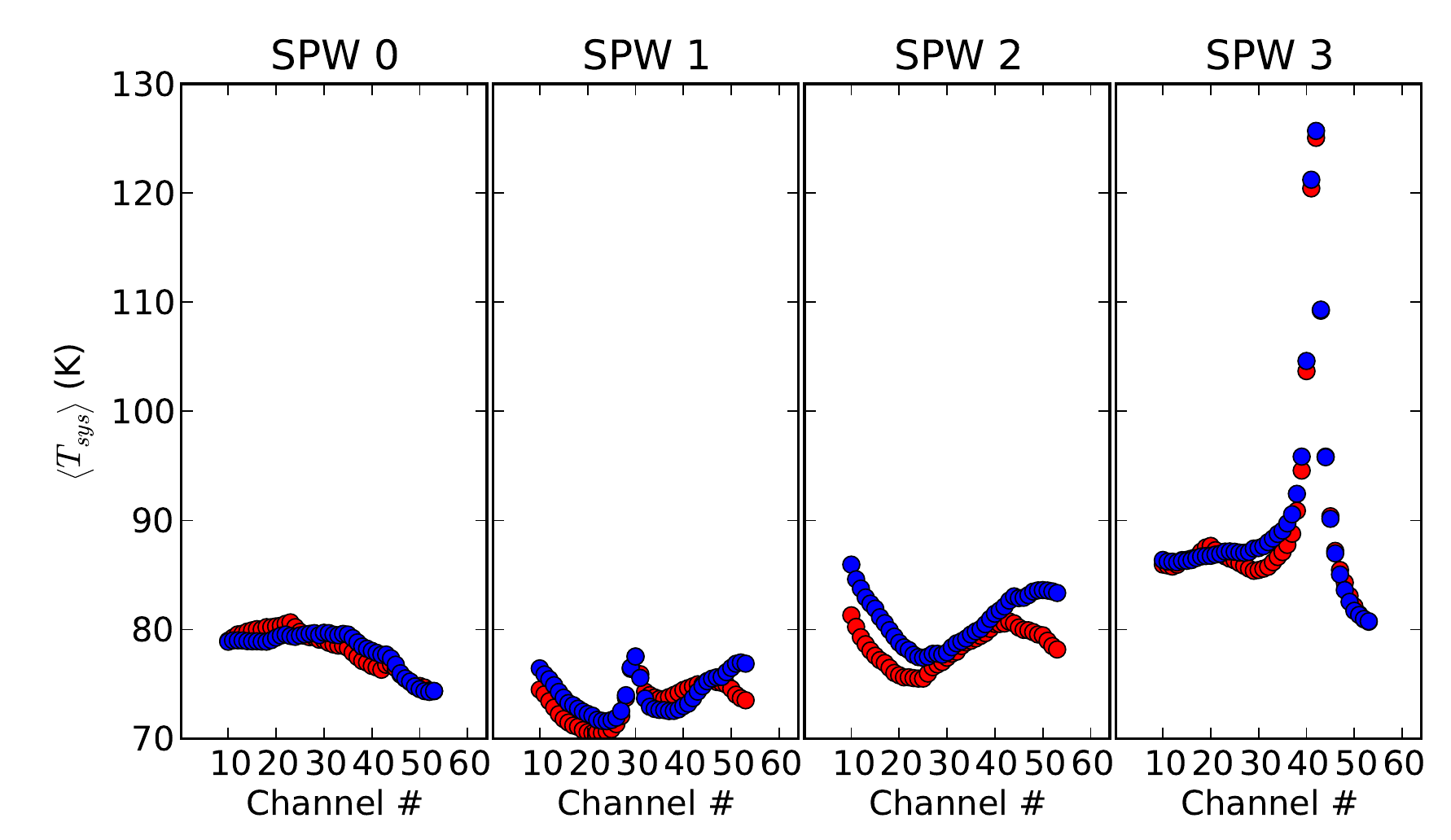}
\caption{System temperature T$_\mathrm{sys}$ as a function of spectral window
  channel for all the spectral windows.}
\label{Tsys}
\end{figure}

Based on the standard calibration and the quality assurance products
provided by the ALMA Arc node, we detect an inconsistency between spw 3
and the rest of the frequency windows, with a discontinuity in
fractional polarization and EVPA that is very likely of instrumental
origin. This window is the most affected by atmospheric opacity, as
shown in Fig.~\ref{Tsys}, which is why we perform additional
calibration steps, as described in Sect. \ref{sect:obs}.

The accuracy of the ALMA polarimetry, which may be affected by an
imperfect polarization calibration, has been assessed via a Monte
Carlo analysis, following a procedure similar to that explained in
\cite{vlemmings17} (see their Appendix A). These simulations are
  specially designed to account for the systematic uncertainties
  arising from the polarization calibration.

In our Monte Carlo analysis, each gain contribution to the full
calibration (i.e., bandpass, amplitude and phase gains, polarization
leakage, amplitude ratio between polarizers, and phase spectrum
between the polarizers at the reference antenna) was perturbed with
random noise. For each realization of the noise, the complete set of
Stokes parameters of the target was derived by fitting a point source
located at the phase center. This translated into a distribution of
Stokes parameters, which reflect the posterior probability density of
the target parameters, as given from the assumed noise distribution of
the antenna gains. The parameters of the gain-noise distributions are
summarized in Table \ref{MCTable}. In addition to the noise added to the calibration tables, an extra contribution of thermal noise was added to each Stokes parameter at each iteration to reflect the effects of the finite ALMA sensitivity. 

The noise parameters given in Table \ref{MCTable} are based on an
educated guess, from our experience with ALMA polarimetry calibration
(with the exception of the baseline sensitivity, which is taken from
the thermal noise at ALMA, as estimated \red{from the residuals of the
  Stokes V image, which is free of dynamic range limitations}). The gains of each spectral window were perturbed with independent gain noises, with the exception of the X-Y phase (which is fitted by CASA using the same polarization model for all spws).

\begin{table}
\begin{tabular}{lll}
\hline
\hline
Noise type               &  Noise mean &  Noise std. \\
\hline
D-terms (real and imag)   &      0      &     1\%   \\  
X-Y Phase                &      0      &     2\,deg. \\ 
X-Y bandpass (amplitude) &      0      &     0.1\%   \\
X-Y bandpass (phase)     &      0      &   0.5\,deg. \\
Thermal (cont.)          &      0      &     \red{8.5$\times 10^{-5}$\,Jy}
\\
\hline
\end{tabular}
\caption{Parameters of the Gaussian noise distributions used to
  perturb the ALMA data in our Monte Carlo polarimetry
  assessment. ``X-Y Phase'' is the residual phase between the
  polarizers at the reference antenna. ``X-Y bandpass'' is the
  channel-wise residual noise of the bandpass, relative between
  polarizers. ``Thermal'' is the theoretical image sensitivity in the
  continuum (as estimated from the \red{Stokes V image}). ``D-terms'' is the channel-wise residual polarization leakage.}
\label{MCTable}
\end{table}

\begin{figure}
\includegraphics[width=\hsize]{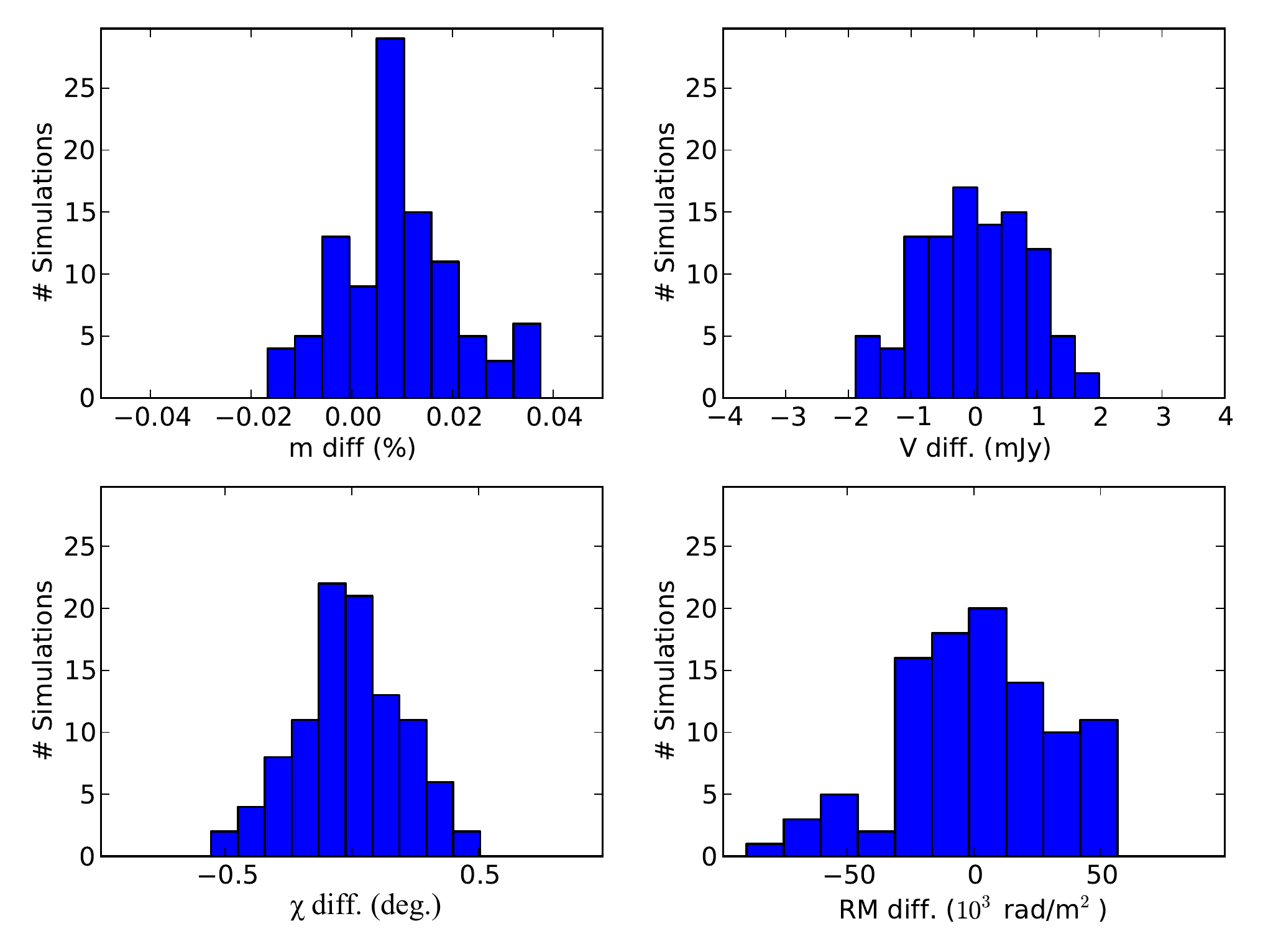}
\caption{Distributions of the difference between polarization
  properties \red{and the RM} from the Monte Carlo iterations and those derived from the calibration.}
\label{PolMCFig}
\end{figure}

\subsection{Monte Carlo results I: linear polarization}

We show in Fig. \ref{PolMCFig} the distributions of the differences
(averaged over all frequency channels) between the polarization
properties derived from the Monte Carlo iterations and the
polarization properties derived from the original ALMA calibration
tables. The accuracy in the estimate of the fractional linear
polarization (which we estimate from the dispersion of $m$
differences) is very high, on the order of 0.02\%. The gain-noise
contribution to the polarization angle, $\chi$, is on the order of
\red{0.5\degr}.

We have also computed the rotation measure (RM) from each Monte Carlo
iteration. The distribution of $RM$ deviations with respect to the
value estimated from the original calibration is \red{also} shown in
Fig. \ref{PolMCFig}. The standard deviation is
\red{$\sim3\times10^4$~\rad, consistent with the uncertainty
  estimate from the Faraday rotation synthesis and simple qu-fitting procedures}. 

\subsection{Monte Carlo results II: circular polarization}

The uncertainty in Stokes $V$, as estimated from Fig. \ref{PolMCFig} (standard deviation of \red{0.9\,mJy}), is a large fraction of the total Stokes $V$ derived from the original calibration (continuum average of $\sim$10\,mJy). This indicates that part of the detected $V$ may be due to gain noise (in particular to inaccuracies in the estimate of the cross-polarization phases at the reference antenna). However, we note that any spurious Stokes $V$ caused by this kind of calibration errors would depend on the source parallactic angle since the spurious $V$ would be related to leakage from the linear polarization in the frame solidary to the antenna mounts. In particular, a calibration-related spurious circular polarization, $V^\mathrm{sp}$, would be given by

\red{
\begin{equation}
V^{\mathrm{sp}} = V^{\mathrm{true}}\,cos{\Delta} + I\,m\,\sin{\left(2[\chi - \psi]\right)}\,\sin{\Delta}
\label{VspEq}
\end{equation}
}

\noindent \red{where $\Delta$ is an uncalibrated
instrumental phase offset between the $X$ and $Y$ polarizers of the
reference antenna, $\chi$ is the EVPA of the source, $\psi$ is the feed
angle (i.e., parallactic angle plus rotation of the band 6 feed in the
antenna frame\footnote{\red{This angle is $-45^{\circ}$ for ALMA, according
to the metadata provided with the observations.}}), $V^{\mathrm{true}}$ is
the Stokes $V$ intrinsic to the source, $I$ is the total source
intensity and $m$ is the fractional linear polarization. }

\begin{figure}
\includegraphics[width=\hsize]{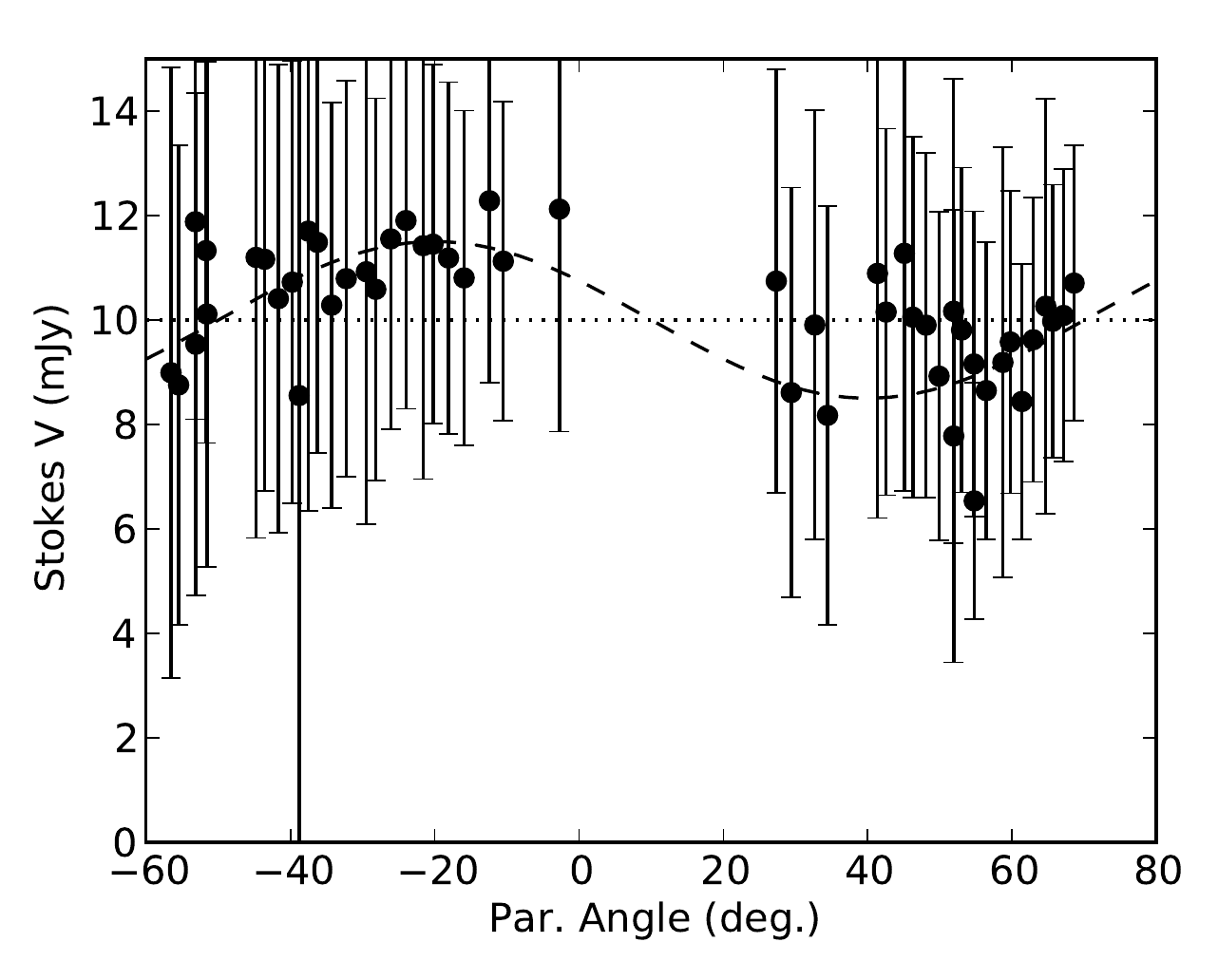}
\caption{Frequency-averaged Stokes $V$, recovered from the original
  ALMA calibration, as a function of parallactic angle. \red{The
    sinusoidal curve corresponds to Eq.~\ref{VspEq}.}}
\label{VTimeFig}
\end{figure}

In Fig. \ref{VTimeFig}, we show the Stokes $V$ recovered from the
original ALMA calibration (averaged over all the spectral channels) as
a function of parallactic angle. A sinusoidal dependence of $V$ is
seen as a function of parallactic angle, which cannot be related to
emission intrinsic to the source. This dependence can be explained as
being due to calibration artifacts (Eq. \ref{VspEq}), i.e., introduced
by leakage from the linear polarization due to an error in the
cross-polarization phase, $\Delta$. 

\red{Indeed, given that the average EVPA is $\chi \sim -60^{\circ}$,
we would expect the sinusoid to have a null close to a parallactic angle
$\sim (-60 + 45)^{\circ} \sim 15^{\circ}$ (i.e., we when the EVPA
is parallel to the $Y$ axis of the polarizers, hence resulting in a
null $U$ Stokes in the antenna frame). This seems to be the case,
according to Fig. \ref{VTimeFig}.}

Using Eq. \ref{VspEq}, we estimate that the observed dependence of $V$ with parallactic angle can be produced by a cross-polarization phase offset of \red{$\Delta \sim 1.5$\degr}, given a fractional linear polarization of \red{$\sim$1.8\%}. The uncertainty level in Stokes $V$ derived from our Monte Carlo simulations (Fig. \ref{PolMCFig}) is indeed on the order of the amplitude in the sinusoid seen in the calibrated data (Fig. \ref{VTimeFig}), which indicates that our Monte Carlo estimates of polarimetric uncertainties are realistic.

We notice, though, that there is another contribution to Stokes $V$ (at a level of about 10\,mJy) that is independent of the parallactic angle and may be related to intrinsic Stokes $V$ from the source (i.e., $V^\mathrm{true}$ in Eq. \ref{VspEq}). This conclusion should, however, be taken with care. Circular polarization is still not officially supported by the ALMA observatory, and even though we do account for all the residual gain factors in our Monte Carlo analysis (see Table \ref{MCTable}), we would need further ALMA observations (with a properly supported circular polarimetry) for the re-assessment of these results.

In short, the conclusion from the assessment of the circular polarization is that there is evidence of spurious contribution from gain noise (via the dependence of $V$ with parallactic angle), but also evidence of intrinsic Stokes $V$ from the source. In any case, the low level of Stokes $V$, together with other contributions from the gain noise, make the flux-density estimate of such an intrinsic $V$ component difficult.

\end{appendix}

\Online
\onllongtab{
\begin{longtable}{lllllll} 
\caption{Stokes parameters and the derived polarization fraction and EVPA for each spectral channel.}\\ 
\hline
\hline
Freq. & Stokes I & Stokes Q & Stokes U & Stokes V & Pol. fraction & EVPA \\ 
(GHz) & (Jy) & (Jy) & (Jy) & (Jy) & (\%) & (\degr) \\ 
\hline
\endfirsthead 
\caption{continued.}\\ 
\hline
\hline
Freq. & Stokes I & Stokes Q & Stokes U & Stokes V & Pol. fraction & EVPA \\ 
(GHz) & (Jy) & (Jy) & (Jy) & (Jy) & (\%) & (\degr) \\ 
\hline
\endhead 
\hline
\endfoot
223.17 & 4.790$\pm$0.009 & $-0.039 \pm 0.007$ & -0.090$\pm$0.009 & 0.010$\pm$0.013 & 2.05$\pm$0.19 & -56.60$\pm$1.99\\ 
223.20 & 4.790$\pm$0.008 & $-0.039 \pm 0.008$ & -0.089$\pm$0.008 & 0.012$\pm$0.013 & 2.03$\pm$0.16 & -56.69$\pm$2.44\\ 
223.23 & 4.789$\pm$0.009 & $-0.039 \pm 0.009$ & -0.090$\pm$0.008 & 0.012$\pm$0.014 & 2.05$\pm$0.16 & -56.55$\pm$2.49\\ 
223.27 & 4.789$\pm$0.009 & $-0.039 \pm 0.009$ & -0.093$\pm$0.008 & 0.011$\pm$0.014 & 2.10$\pm$0.18 & -56.28$\pm$2.22\\ 
223.30 & 4.788$\pm$0.008 & $-0.039 \pm 0.008$ & -0.094$\pm$0.008 & 0.010$\pm$0.013 & 2.12$\pm$0.17 & -56.17$\pm$2.33\\ 
223.33 & 4.788$\pm$0.009 & $-0.039 \pm 0.008$ & -0.093$\pm$0.007 & 0.009$\pm$0.013 & 2.10$\pm$0.16 & -56.28$\pm$2.24\\ 
223.36 & 4.787$\pm$0.009 & $-0.039 \pm 0.010$ & -0.091$\pm$0.007 & 0.010$\pm$0.013 & 2.06$\pm$0.16 & -56.51$\pm$2.73\\ 
223.39 & 4.787$\pm$0.008 & $-0.039 \pm 0.009$ & -0.091$\pm$0.009 & 0.010$\pm$0.012 & 2.07$\pm$0.19 & -56.44$\pm$2.43\\ 
223.42 & 4.786$\pm$0.008 & $-0.039 \pm 0.009$ & -0.093$\pm$0.008 & 0.008$\pm$0.013 & 2.10$\pm$0.16 & -56.28$\pm$2.62\\ 
223.45 & 4.786$\pm$0.008 & $-0.039 \pm 0.010$ & -0.093$\pm$0.009 & 0.009$\pm$0.012 & 2.10$\pm$0.18 & -56.29$\pm$2.83\\ 
223.48 & 4.785$\pm$0.009 & $-0.039 \pm 0.009$ & -0.093$\pm$0.008 & 0.010$\pm$0.012 & 2.11$\pm$0.17 & -56.24$\pm$2.71\\ 
223.52 & 4.785$\pm$0.008 & $-0.039 \pm 0.009$ & -0.092$\pm$0.009 & 0.010$\pm$0.011 & 2.09$\pm$0.19 & -56.33$\pm$2.42\\ 
223.55 & 4.784$\pm$0.009 & $-0.039 \pm 0.008$ & -0.089$\pm$0.009 & 0.011$\pm$0.014 & 2.03$\pm$0.19 & -56.67$\pm$2.47\\ 
223.58 & 4.784$\pm$0.007 & $-0.039 \pm 0.008$ & -0.087$\pm$0.008 & 0.011$\pm$0.014 & 2.00$\pm$0.16 & -56.90$\pm$2.29\\ 
223.61 & 4.783$\pm$0.009 & $-0.039 \pm 0.008$ & -0.090$\pm$0.008 & 0.008$\pm$0.013 & 2.04$\pm$0.16 & -56.64$\pm$2.45\\ 
223.64 & 4.783$\pm$0.008 & $-0.039 \pm 0.008$ & -0.091$\pm$0.008 & 0.008$\pm$0.014 & 2.07$\pm$0.18 & -56.45$\pm$2.18\\ 
223.67 & 4.782$\pm$0.008 & $-0.039 \pm 0.009$ & -0.090$\pm$0.009 & 0.008$\pm$0.013 & 2.05$\pm$0.19 & -56.57$\pm$2.48\\ 
223.70 & 4.782$\pm$0.009 & $-0.039 \pm 0.008$ & -0.091$\pm$0.008 & 0.008$\pm$0.013 & 2.07$\pm$0.16 & -56.43$\pm$2.46\\ 
223.73 & 4.781$\pm$0.009 & $-0.039 \pm 0.007$ & -0.094$\pm$0.008 & 0.008$\pm$0.012 & 2.13$\pm$0.16 & -56.11$\pm$2.13\\ 
223.77 & 4.781$\pm$0.008 & $-0.039 \pm 0.008$ & -0.093$\pm$0.008 & 0.008$\pm$0.012 & 2.11$\pm$0.17 & -56.22$\pm$2.05\\ 
223.80 & 4.780$\pm$0.009 & $-0.039 \pm 0.008$ & -0.091$\pm$0.008 & 0.009$\pm$0.012 & 2.06$\pm$0.17 & -56.51$\pm$2.46\\ 
223.83 & 4.780$\pm$0.007 & $-0.038 \pm 0.007$ & -0.090$\pm$0.008 & 0.010$\pm$0.013 & 2.05$\pm$0.16 & -56.54$\pm$2.25\\ 
223.86 & 4.779$\pm$0.010 & $-0.038 \pm 0.007$ & -0.090$\pm$0.009 & 0.009$\pm$0.013 & 2.05$\pm$0.19 & -56.57$\pm$2.23\\ 
223.89 & 4.779$\pm$0.008 & $-0.038 \pm 0.008$ & -0.091$\pm$0.009 & 0.008$\pm$0.014 & 2.06$\pm$0.18 & -56.51$\pm$2.38\\ 
223.92 & 4.778$\pm$0.009 & $-0.038 \pm 0.009$ & -0.090$\pm$0.008 & 0.008$\pm$0.013 & 2.04$\pm$0.15 & -56.62$\pm$2.59\\ 
223.95 & 4.778$\pm$0.008 & $-0.038 \pm 0.007$ & -0.088$\pm$0.008 & 0.008$\pm$0.013 & 2.02$\pm$0.17 & -56.75$\pm$2.21\\ 
223.98 & 4.777$\pm$0.007 & $-0.038 \pm 0.009$ & -0.090$\pm$0.008 & 0.008$\pm$0.013 & 2.05$\pm$0.16 & -56.54$\pm$2.66\\ 
224.02 & 4.777$\pm$0.008 & $-0.038 \pm 0.009$ & -0.090$\pm$0.008 & 0.008$\pm$0.012 & 2.04$\pm$0.17 & -56.61$\pm$2.56\\ 
224.05 & 4.776$\pm$0.009 & $-0.038 \pm 0.008$ & -0.088$\pm$0.009 & 0.010$\pm$0.015 & 2.01$\pm$0.18 & -56.83$\pm$2.22\\ 
224.08 & 4.776$\pm$0.008 & $-0.038 \pm 0.007$ & -0.088$\pm$0.009 & 0.011$\pm$0.014 & 2.01$\pm$0.19 & -56.79$\pm$2.21\\ 
224.11 & 4.775$\pm$0.007 & $-0.038 \pm 0.007$ & -0.088$\pm$0.008 & 0.011$\pm$0.012 & 2.00$\pm$0.17 & -56.86$\pm$2.18\\ 
224.14 & 4.775$\pm$0.009 & $-0.038 \pm 0.007$ & -0.088$\pm$0.008 & 0.011$\pm$0.012 & 2.00$\pm$0.16 & -56.86$\pm$2.28\\ 
224.17 & 4.774$\pm$0.008 & $-0.038 \pm 0.008$ & -0.089$\pm$0.008 & 0.010$\pm$0.014 & 2.03$\pm$0.16 & -56.68$\pm$2.22\\ 
224.20 & 4.774$\pm$0.007 & $-0.038 \pm 0.008$ & -0.091$\pm$0.008 & 0.008$\pm$0.011 & 2.07$\pm$0.16 & -56.45$\pm$2.28\\ 
224.23 & 4.773$\pm$0.008 & $-0.038 \pm 0.009$ & -0.093$\pm$0.008 & 0.009$\pm$0.013 & 2.10$\pm$0.18 & -56.25$\pm$2.31\\ 
224.27 & 4.773$\pm$0.009 & $-0.038 \pm 0.008$ & -0.093$\pm$0.009 & 0.009$\pm$0.015 & 2.11$\pm$0.19 & -56.23$\pm$2.32\\ 
224.30 & 4.772$\pm$0.008 & $-0.038 \pm 0.008$ & -0.093$\pm$0.008 & 0.009$\pm$0.012 & 2.11$\pm$0.17 & -56.22$\pm$2.12\\ 
224.33 & 4.772$\pm$0.009 & $-0.038 \pm 0.008$ & -0.093$\pm$0.009 & 0.010$\pm$0.013 & 2.10$\pm$0.20 & -56.25$\pm$2.11\\ 
224.36 & 4.771$\pm$0.008 & $-0.038 \pm 0.008$ & -0.091$\pm$0.008 & 0.012$\pm$0.012 & 2.07$\pm$0.17 & -56.45$\pm$2.41\\ 
224.39 & 4.771$\pm$0.009 & $-0.038 \pm 0.008$ & -0.089$\pm$0.009 & 0.012$\pm$0.013 & 2.04$\pm$0.19 & -56.62$\pm$2.32\\ 
224.42 & 4.770$\pm$0.009 & $-0.038 \pm 0.008$ & -0.089$\pm$0.008 & 0.010$\pm$0.014 & 2.04$\pm$0.16 & -56.66$\pm$2.33\\ 
224.45 & 4.770$\pm$0.007 & $-0.038 \pm 0.008$ & -0.090$\pm$0.008 & 0.010$\pm$0.013 & 2.06$\pm$0.17 & -56.53$\pm$2.37\\ 
224.48 & 4.769$\pm$0.008 & $-0.038 \pm 0.008$ & -0.090$\pm$0.009 & 0.011$\pm$0.012 & 2.06$\pm$0.18 & -56.50$\pm$2.35\\ 
224.52 & 4.769$\pm$0.009 & $-0.038 \pm 0.008$ & -0.089$\pm$0.009 & 0.009$\pm$0.012 & 2.03$\pm$0.18 & -56.66$\pm$2.41\\ 
224.55 & 4.768$\pm$0.009 & $-0.038 \pm 0.008$ & -0.089$\pm$0.008 & 0.007$\pm$0.013 & 2.03$\pm$0.17 & -56.71$\pm$2.32\\ 
224.58 & 4.768$\pm$0.009 & $-0.038 \pm 0.008$ & -0.090$\pm$0.009 & 0.008$\pm$0.014 & 2.04$\pm$0.19 & -56.60$\pm$2.41\\ 
224.61 & 4.767$\pm$0.008 & $-0.038 \pm 0.008$ & -0.090$\pm$0.008 & 0.009$\pm$0.013 & 2.05$\pm$0.17 & -56.56$\pm$2.47\\ 
224.64 & 4.767$\pm$0.007 & $-0.038 \pm 0.007$ & -0.090$\pm$0.008 & 0.011$\pm$0.013 & 2.06$\pm$0.16 & -56.50$\pm$2.30\\ 
224.67 & 4.766$\pm$0.009 & $-0.038 \pm 0.009$ & -0.092$\pm$0.008 & 0.011$\pm$0.011 & 2.09$\pm$0.18 & -56.30$\pm$2.55\\ 
224.70 & 4.766$\pm$0.008 & $-0.038 \pm 0.008$ & -0.090$\pm$0.008 & 0.010$\pm$0.013 & 2.06$\pm$0.18 & -56.51$\pm$2.45\\ 
224.73 & 4.765$\pm$0.009 & $-0.038 \pm 0.007$ & -0.088$\pm$0.008 & 0.009$\pm$0.013 & 2.01$\pm$0.17 & -56.84$\pm$2.23\\ 
224.77 & 4.765$\pm$0.008 & $-0.038 \pm 0.008$ & -0.089$\pm$0.009 & 0.009$\pm$0.012 & 2.03$\pm$0.19 & -56.72$\pm$2.27\\ 
224.80 & 4.764$\pm$0.008 & $-0.038 \pm 0.008$ & -0.088$\pm$0.009 & 0.010$\pm$0.013 & 2.02$\pm$0.19 & -56.74$\pm$2.40\\ 
224.83 & 4.764$\pm$0.008 & $-0.038 \pm 0.007$ & -0.086$\pm$0.007 & 0.011$\pm$0.014 & 1.98$\pm$0.14 & -56.99$\pm$2.18\\ 
225.17 & 4.753$\pm$0.009 & $-0.039 \pm 0.008$ & -0.088$\pm$0.008 & 0.010$\pm$0.016 & 2.02$\pm$0.18 & -56.87$\pm$2.41\\ 
225.20 & 4.752$\pm$0.009 & $-0.039 \pm 0.007$ & -0.088$\pm$0.008 & 0.011$\pm$0.012 & 2.01$\pm$0.16 & -56.91$\pm$2.17\\ 
225.23 & 4.752$\pm$0.009 & $-0.039 \pm 0.008$ & -0.087$\pm$0.009 & 0.011$\pm$0.014 & 2.00$\pm$0.17 & -57.03$\pm$2.59\\ 
225.27 & 4.751$\pm$0.009 & $-0.039 \pm 0.009$ & -0.086$\pm$0.008 & 0.012$\pm$0.013 & 1.99$\pm$0.15 & -57.04$\pm$2.72\\ 
225.30 & 4.751$\pm$0.009 & $-0.039 \pm 0.009$ & -0.087$\pm$0.009 & 0.011$\pm$0.014 & 2.00$\pm$0.18 & -56.99$\pm$2.72\\ 
225.33 & 4.750$\pm$0.008 & $-0.039 \pm 0.009$ & -0.088$\pm$0.008 & 0.010$\pm$0.012 & 2.02$\pm$0.18 & -56.89$\pm$2.57\\ 
225.36 & 4.750$\pm$0.008 & $-0.039 \pm 0.009$ & -0.088$\pm$0.009 & 0.010$\pm$0.014 & 2.02$\pm$0.19 & -56.85$\pm$2.59\\ 
225.39 & 4.749$\pm$0.008 & $-0.039 \pm 0.008$ & -0.087$\pm$0.007 & 0.010$\pm$0.013 & 2.00$\pm$0.16 & -56.99$\pm$2.37\\ 
225.42 & 4.749$\pm$0.008 & $-0.039 \pm 0.008$ & -0.086$\pm$0.008 & 0.009$\pm$0.013 & 1.99$\pm$0.18 & -57.06$\pm$2.37\\ 
225.45 & 4.748$\pm$0.008 & $-0.039 \pm 0.008$ & -0.088$\pm$0.008 & 0.009$\pm$0.013 & 2.02$\pm$0.15 & -56.89$\pm$2.41\\ 
225.48 & 4.748$\pm$0.008 & $-0.039 \pm 0.008$ & -0.089$\pm$0.009 & 0.008$\pm$0.013 & 2.04$\pm$0.19 & -56.75$\pm$2.33\\ 
225.52 & 4.747$\pm$0.008 & $-0.039 \pm 0.008$ & -0.088$\pm$0.008 & 0.009$\pm$0.014 & 2.02$\pm$0.17 & -56.85$\pm$2.49\\ 
225.55 & 4.747$\pm$0.009 & $-0.039 \pm 0.008$ & -0.086$\pm$0.008 & 0.010$\pm$0.014 & 2.00$\pm$0.16 & -57.03$\pm$2.74\\ 
225.58 & 4.746$\pm$0.008 & $-0.039 \pm 0.008$ & -0.084$\pm$0.008 & 0.011$\pm$0.012 & 1.95$\pm$0.17 & -57.31$\pm$2.49\\ 
225.61 & 4.746$\pm$0.009 & $-0.039 \pm 0.007$ & -0.084$\pm$0.008 & 0.011$\pm$0.013 & 1.95$\pm$0.17 & -57.32$\pm$2.25\\ 
225.64 & 4.745$\pm$0.009 & $-0.039 \pm 0.008$ & -0.086$\pm$0.008 & 0.009$\pm$0.014 & 1.99$\pm$0.17 & -57.09$\pm$2.59\\ 
225.67 & 4.745$\pm$0.009 & $-0.039 \pm 0.007$ & -0.086$\pm$0.007 & 0.007$\pm$0.013 & 1.98$\pm$0.14 & -57.12$\pm$2.38\\ 
225.70 & 4.744$\pm$0.009 & $-0.039 \pm 0.008$ & -0.086$\pm$0.008 & 0.007$\pm$0.014 & 1.99$\pm$0.16 & -57.05$\pm$2.54\\ 
225.73 & 4.744$\pm$0.008 & $-0.039 \pm 0.008$ & -0.088$\pm$0.010 & 0.008$\pm$0.012 & 2.03$\pm$0.20 & -56.79$\pm$2.50\\ 
225.77 & 4.743$\pm$0.008 & $-0.039 \pm 0.009$ & -0.088$\pm$0.008 & 0.009$\pm$0.013 & 2.02$\pm$0.19 & -56.89$\pm$2.56\\ 
225.80 & 4.743$\pm$0.009 & $-0.039 \pm 0.009$ & -0.085$\pm$0.008 & 0.008$\pm$0.013 & 1.97$\pm$0.18 & -57.19$\pm$2.50\\ 
225.83 & 4.742$\pm$0.009 & $-0.039 \pm 0.008$ & -0.084$\pm$0.009 & 0.009$\pm$0.013 & 1.95$\pm$0.18 & -57.30$\pm$2.73\\ 
225.86 & 4.742$\pm$0.008 & $-0.039 \pm 0.008$ & -0.084$\pm$0.008 & 0.010$\pm$0.014 & 1.95$\pm$0.16 & -57.33$\pm$2.54\\ 
225.89 & 4.741$\pm$0.008 & $-0.039 \pm 0.009$ & -0.082$\pm$0.009 & 0.010$\pm$0.015 & 1.92$\pm$0.20 & -57.53$\pm$2.65\\ 
225.92 & 4.741$\pm$0.009 & $-0.039 \pm 0.009$ & -0.082$\pm$0.008 & 0.009$\pm$0.012 & 1.91$\pm$0.18 & -57.58$\pm$2.60\\ 
225.95 & 4.740$\pm$0.008 & $-0.039 \pm 0.008$ & -0.085$\pm$0.008 & 0.008$\pm$0.012 & 1.97$\pm$0.16 & -57.18$\pm$2.60\\ 
225.98 & 4.740$\pm$0.008 & $-0.039 \pm 0.008$ & -0.086$\pm$0.008 & 0.007$\pm$0.013 & 1.99$\pm$0.15 & -57.04$\pm$2.58\\ 
226.02 & 4.739$\pm$0.008 & $-0.039 \pm 0.008$ & -0.085$\pm$0.008 & 0.007$\pm$0.015 & 1.97$\pm$0.16 & -57.21$\pm$2.57\\ 
226.05 & 4.739$\pm$0.008 & $-0.039 \pm 0.008$ & -0.084$\pm$0.008 & 0.007$\pm$0.015 & 1.95$\pm$0.17 & -57.30$\pm$2.61\\ 
226.08 & 4.738$\pm$0.008 & $-0.039 \pm 0.008$ & -0.084$\pm$0.009 & 0.009$\pm$0.013 & 1.94$\pm$0.18 & -57.37$\pm$2.55\\ 
226.11 & 4.738$\pm$0.009 & $-0.039 \pm 0.009$ & -0.084$\pm$0.008 & 0.009$\pm$0.014 & 1.95$\pm$0.17 & -57.35$\pm$2.70\\ 
226.14 & 4.737$\pm$0.008 & $-0.039 \pm 0.009$ & -0.084$\pm$0.008 & 0.009$\pm$0.012 & 1.96$\pm$0.19 & -57.27$\pm$2.73\\ 
226.17 & 4.737$\pm$0.008 & $-0.039 \pm 0.009$ & -0.085$\pm$0.008 & 0.009$\pm$0.013 & 1.96$\pm$0.16 & -57.23$\pm$2.54\\ 
226.20 & 4.736$\pm$0.009 & $-0.039 \pm 0.008$ & -0.084$\pm$0.008 & 0.009$\pm$0.012 & 1.96$\pm$0.17 & -57.29$\pm$2.65\\ 
226.23 & 4.736$\pm$0.009 & $-0.039 \pm 0.008$ & -0.084$\pm$0.009 & 0.009$\pm$0.012 & 1.94$\pm$0.19 & -57.36$\pm$2.47\\ 
226.27 & 4.735$\pm$0.008 & $-0.039 \pm 0.008$ & -0.085$\pm$0.008 & 0.009$\pm$0.014 & 1.98$\pm$0.18 & -57.16$\pm$2.49\\ 
226.30 & 4.735$\pm$0.008 & $-0.039 \pm 0.008$ & -0.085$\pm$0.007 & 0.009$\pm$0.012 & 1.97$\pm$0.16 & -57.19$\pm$2.50\\ 
226.33 & 4.734$\pm$0.008 & $-0.039 \pm 0.008$ & -0.082$\pm$0.010 & 0.009$\pm$0.015 & 1.91$\pm$0.19 & -57.58$\pm$2.97\\ 
226.36 & 4.734$\pm$0.007 & $-0.039 \pm 0.008$ & -0.081$\pm$0.009 & 0.010$\pm$0.013 & 1.90$\pm$0.19 & -57.67$\pm$2.56\\ 
226.39 & 4.733$\pm$0.009 & $-0.039 \pm 0.009$ & -0.083$\pm$0.009 & 0.009$\pm$0.012 & 1.92$\pm$0.19 & -57.50$\pm$2.80\\ 
226.42 & 4.733$\pm$0.009 & $-0.039 \pm 0.007$ & -0.083$\pm$0.007 & 0.009$\pm$0.013 & 1.94$\pm$0.15 & -57.40$\pm$2.20\\ 
226.45 & 4.732$\pm$0.008 & $-0.039 \pm 0.007$ & -0.085$\pm$0.007 & 0.009$\pm$0.013 & 1.97$\pm$0.16 & -57.19$\pm$2.17\\ 
226.48 & 4.732$\pm$0.009 & $-0.038 \pm 0.008$ & -0.086$\pm$0.007 & 0.010$\pm$0.014 & 1.99$\pm$0.16 & -57.06$\pm$2.29\\ 
226.52 & 4.731$\pm$0.007 & $-0.038 \pm 0.008$ & -0.084$\pm$0.009 & 0.011$\pm$0.013 & 1.96$\pm$0.19 & -57.29$\pm$2.50\\ 
226.55 & 4.731$\pm$0.008 & $-0.038 \pm 0.009$ & -0.083$\pm$0.008 & 0.012$\pm$0.015 & 1.93$\pm$0.16 & -57.48$\pm$2.73\\ 
226.58 & 4.730$\pm$0.009 & $-0.038 \pm 0.009$ & -0.083$\pm$0.007 & 0.011$\pm$0.016 & 1.93$\pm$0.15 & -57.48$\pm$2.71\\ 
226.61 & 4.730$\pm$0.009 & $-0.038 \pm 0.008$ & -0.083$\pm$0.009 & 0.009$\pm$0.014 & 1.92$\pm$0.19 & -57.50$\pm$2.46\\ 
226.64 & 4.729$\pm$0.008 & $-0.038 \pm 0.008$ & -0.081$\pm$0.008 & 0.009$\pm$0.012 & 1.90$\pm$0.17 & -57.71$\pm$2.52\\ 
226.67 & 4.729$\pm$0.008 & $-0.038 \pm 0.008$ & -0.079$\pm$0.008 & 0.012$\pm$0.014 & 1.87$\pm$0.16 & -57.91$\pm$2.92\\ 
226.70 & 4.728$\pm$0.009 & $-0.038 \pm 0.008$ & -0.081$\pm$0.009 & 0.012$\pm$0.014 & 1.90$\pm$0.19 & -57.68$\pm$2.57\\ 
226.73 & 4.728$\pm$0.008 & $-0.038 \pm 0.008$ & -0.084$\pm$0.007 & 0.011$\pm$0.013 & 1.95$\pm$0.17 & -57.32$\pm$2.47\\ 
226.77 & 4.727$\pm$0.008 & $-0.038 \pm 0.008$ & -0.083$\pm$0.008 & 0.011$\pm$0.014 & 1.93$\pm$0.18 & -57.44$\pm$2.48\\ 
226.80 & 4.727$\pm$0.008 & $-0.038 \pm 0.009$ & -0.081$\pm$0.008 & 0.011$\pm$0.012 & 1.90$\pm$0.17 & -57.71$\pm$2.71\\ 
226.83 & 4.726$\pm$0.008 & $-0.038 \pm 0.008$ & -0.082$\pm$0.008 & 0.011$\pm$0.014 & 1.91$\pm$0.18 & -57.59$\pm$2.49\\ 
239.17 & 4.495$\pm$0.009 & $-0.043 \pm 0.009$ & -0.060$\pm$0.009 & 0.011$\pm$0.011 & 1.64$\pm$0.19 & -62.66$\pm$3.33\\ 
239.20 & 4.495$\pm$0.008 & $-0.043 \pm 0.008$ & -0.059$\pm$0.008 & 0.012$\pm$0.012 & 1.61$\pm$0.17 & -63.03$\pm$3.31\\ 
239.23 & 4.494$\pm$0.008 & $-0.043 \pm 0.009$ & -0.058$\pm$0.009 & 0.010$\pm$0.011 & 1.61$\pm$0.21 & -63.07$\pm$3.17\\ 
239.27 & 4.494$\pm$0.008 & $-0.043 \pm 0.008$ & -0.060$\pm$0.008 & 0.010$\pm$0.010 & 1.64$\pm$0.16 & -62.68$\pm$3.33\\ 
239.30 & 4.493$\pm$0.008 & $-0.043 \pm 0.008$ & -0.062$\pm$0.008 & 0.010$\pm$0.014 & 1.67$\pm$0.18 & -62.37$\pm$3.11\\ 
239.33 & 4.493$\pm$0.009 & $-0.043 \pm 0.009$ & -0.060$\pm$0.008 & 0.009$\pm$0.013 & 1.63$\pm$0.19 & -62.81$\pm$3.36\\ 
239.36 & 4.493$\pm$0.009 & $-0.043 \pm 0.009$ & -0.057$\pm$0.009 & 0.010$\pm$0.014 & 1.59$\pm$0.17 & -63.39$\pm$3.90\\ 
239.39 & 4.492$\pm$0.008 & $-0.043 \pm 0.008$ & -0.058$\pm$0.009 & 0.010$\pm$0.011 & 1.61$\pm$0.21 & -63.09$\pm$2.97\\ 
239.42 & 4.492$\pm$0.009 & $-0.043 \pm 0.009$ & -0.060$\pm$0.009 & 0.009$\pm$0.013 & 1.64$\pm$0.21 & -62.71$\pm$3.24\\ 
239.45 & 4.491$\pm$0.009 & $-0.043 \pm 0.008$ & -0.060$\pm$0.007 & 0.008$\pm$0.012 & 1.64$\pm$0.16 & -62.72$\pm$3.07\\ 
239.48 & 4.491$\pm$0.009 & $-0.043 \pm 0.007$ & -0.061$\pm$0.007 & 0.009$\pm$0.014 & 1.66$\pm$0.16 & -62.50$\pm$2.96\\ 
239.52 & 4.490$\pm$0.008 & $-0.043 \pm 0.009$ & -0.063$\pm$0.007 & 0.010$\pm$0.011 & 1.70$\pm$0.16 & -62.05$\pm$3.48\\ 
239.55 & 4.490$\pm$0.008 & $-0.043 \pm 0.008$ & -0.062$\pm$0.009 & 0.011$\pm$0.012 & 1.68$\pm$0.19 & -62.23$\pm$3.00\\ 
239.58 & 4.490$\pm$0.008 & $-0.043 \pm 0.008$ & -0.062$\pm$0.009 & 0.011$\pm$0.013 & 1.67$\pm$0.18 & -62.37$\pm$3.32\\ 
239.61 & 4.489$\pm$0.008 & $-0.043 \pm 0.009$ & -0.062$\pm$0.008 & 0.010$\pm$0.015 & 1.68$\pm$0.18 & -62.17$\pm$3.51\\ 
239.64 & 4.489$\pm$0.009 & $-0.043 \pm 0.008$ & -0.061$\pm$0.007 & 0.009$\pm$0.012 & 1.65$\pm$0.16 & -62.54$\pm$3.03\\ 
239.67 & 4.488$\pm$0.008 & $-0.043 \pm 0.008$ & -0.060$\pm$0.007 & 0.011$\pm$0.013 & 1.63$\pm$0.17 & -62.77$\pm$2.73\\ 
239.70 & 4.488$\pm$0.008 & $-0.043 \pm 0.008$ & -0.061$\pm$0.008 & 0.012$\pm$0.013 & 1.66$\pm$0.18 & -62.46$\pm$3.07\\ 
239.73 & 4.487$\pm$0.008 & $-0.043 \pm 0.009$ & -0.062$\pm$0.008 & 0.011$\pm$0.013 & 1.67$\pm$0.20 & -62.29$\pm$3.19\\ 
239.77 & 4.487$\pm$0.009 & $-0.043 \pm 0.008$ & -0.061$\pm$0.007 & 0.010$\pm$0.014 & 1.65$\pm$0.17 & -62.54$\pm$3.02\\ 
239.80 & 4.486$\pm$0.008 & $-0.043 \pm 0.008$ & -0.059$\pm$0.008 & 0.011$\pm$0.011 & 1.63$\pm$0.19 & -62.84$\pm$3.16\\ 
239.83 & 4.486$\pm$0.007 & $-0.043 \pm 0.008$ & -0.059$\pm$0.009 & 0.010$\pm$0.012 & 1.63$\pm$0.19 & -62.83$\pm$3.56\\ 
239.86 & 4.486$\pm$0.009 & $-0.043 \pm 0.008$ & -0.061$\pm$0.008 & 0.010$\pm$0.013 & 1.66$\pm$0.18 & -62.51$\pm$3.07\\ 
239.89 & 4.485$\pm$0.008 & $-0.043 \pm 0.009$ & -0.060$\pm$0.008 & 0.010$\pm$0.013 & 1.65$\pm$0.19 & -62.63$\pm$3.06\\ 
239.92 & 4.485$\pm$0.009 & $-0.043 \pm 0.008$ & -0.058$\pm$0.008 & 0.011$\pm$0.011 & 1.61$\pm$0.19 & -63.15$\pm$2.98\\ 
239.95 & 4.484$\pm$0.010 & $-0.043 \pm 0.008$ & -0.059$\pm$0.009 & 0.011$\pm$0.013 & 1.63$\pm$0.19 & -62.89$\pm$3.21\\ 
239.98 & 4.484$\pm$0.010 & $-0.043 \pm 0.008$ & -0.060$\pm$0.008 & 0.011$\pm$0.013 & 1.65$\pm$0.21 & -62.63$\pm$2.78\\ 
240.02 & 4.483$\pm$0.009 & $-0.043 \pm 0.008$ & -0.058$\pm$0.009 & 0.010$\pm$0.013 & 1.61$\pm$0.19 & -63.03$\pm$3.39\\ 
240.05 & 4.483$\pm$0.009 & $-0.043 \pm 0.009$ & -0.059$\pm$0.007 & 0.010$\pm$0.013 & 1.62$\pm$0.17 & -63.01$\pm$3.45\\ 
240.08 & 4.483$\pm$0.009 & $-0.043 \pm 0.008$ & -0.060$\pm$0.008 & 0.011$\pm$0.012 & 1.65$\pm$0.17 & -62.58$\pm$3.11\\ 
240.11 & 4.482$\pm$0.009 & $-0.043 \pm 0.009$ & -0.061$\pm$0.008 & 0.011$\pm$0.011 & 1.66$\pm$0.18 & -62.51$\pm$3.40\\ 
240.14 & 4.482$\pm$0.009 & $-0.043 \pm 0.007$ & -0.060$\pm$0.008 & 0.010$\pm$0.012 & 1.64$\pm$0.17 & -62.69$\pm$2.66\\ 
240.17 & 4.481$\pm$0.008 & $-0.043 \pm 0.008$ & -0.059$\pm$0.008 & 0.010$\pm$0.013 & 1.63$\pm$0.18 & -62.85$\pm$3.05\\ 
240.20 & 4.481$\pm$0.009 & $-0.043 \pm 0.008$ & -0.060$\pm$0.009 & 0.009$\pm$0.012 & 1.65$\pm$0.21 & -62.58$\pm$3.21\\ 
240.23 & 4.480$\pm$0.007 & $-0.043 \pm 0.008$ & -0.061$\pm$0.007 & 0.010$\pm$0.013 & 1.65$\pm$0.17 & -62.56$\pm$3.16\\ 
240.27 & 4.480$\pm$0.008 & $-0.043 \pm 0.008$ & -0.059$\pm$0.008 & 0.011$\pm$0.012 & 1.62$\pm$0.17 & -63.02$\pm$3.03\\ 
240.30 & 4.479$\pm$0.008 & $-0.043 \pm 0.008$ & -0.057$\pm$0.009 & 0.011$\pm$0.013 & 1.58$\pm$0.19 & -63.49$\pm$3.28\\ 
240.33 & 4.479$\pm$0.008 & $-0.043 \pm 0.008$ & -0.055$\pm$0.009 & 0.011$\pm$0.014 & 1.55$\pm$0.20 & -63.92$\pm$3.10\\ 
240.36 & 4.479$\pm$0.008 & $-0.043 \pm 0.009$ & -0.056$\pm$0.008 & 0.011$\pm$0.013 & 1.58$\pm$0.18 & -63.54$\pm$3.46\\ 
240.39 & 4.478$\pm$0.009 & $-0.043 \pm 0.008$ & -0.058$\pm$0.008 & 0.012$\pm$0.013 & 1.60$\pm$0.18 & -63.21$\pm$3.30\\ 
240.42 & 4.478$\pm$0.009 & $-0.043 \pm 0.009$ & -0.057$\pm$0.008 & 0.011$\pm$0.013 & 1.60$\pm$0.16 & -63.26$\pm$3.71\\ 
240.45 & 4.477$\pm$0.008 & $-0.043 \pm 0.009$ & -0.059$\pm$0.007 & 0.010$\pm$0.013 & 1.62$\pm$0.17 & -62.98$\pm$3.36\\ 
240.48 & 4.477$\pm$0.008 & $-0.043 \pm 0.008$ & -0.056$\pm$0.008 & 0.009$\pm$0.013 & 1.58$\pm$0.18 & -63.54$\pm$3.08\\ 
240.52 & 4.476$\pm$0.008 & $-0.043 \pm 0.008$ & -0.054$\pm$0.009 & 0.009$\pm$0.014 & 1.54$\pm$0.19 & -64.04$\pm$3.53\\ 
240.55 & 4.476$\pm$0.009 & $-0.043 \pm 0.008$ & -0.055$\pm$0.007 & 0.011$\pm$0.013 & 1.56$\pm$0.16 & -63.75$\pm$3.54\\ 
240.58 & 4.476$\pm$0.008 & $-0.043 \pm 0.008$ & -0.055$\pm$0.008 & 0.011$\pm$0.015 & 1.56$\pm$0.18 & -63.83$\pm$3.46\\ 
240.61 & 4.475$\pm$0.009 & $-0.043 \pm 0.008$ & -0.055$\pm$0.009 & 0.011$\pm$0.013 & 1.55$\pm$0.19 & -63.85$\pm$3.61\\ 
240.64 & 4.475$\pm$0.009 & $-0.043 \pm 0.008$ & -0.054$\pm$0.008 & 0.009$\pm$0.012 & 1.53$\pm$0.18 & -64.13$\pm$3.19\\ 
240.67 & 4.474$\pm$0.009 & $-0.043 \pm 0.009$ & -0.053$\pm$0.008 & 0.010$\pm$0.013 & 1.52$\pm$0.17 & -64.40$\pm$3.73\\ 
240.70 & 4.474$\pm$0.008 & $-0.043 \pm 0.009$ & -0.052$\pm$0.008 & 0.011$\pm$0.013 & 1.51$\pm$0.19 & -64.51$\pm$3.91\\ 
240.73 & 4.473$\pm$0.008 & $-0.043 \pm 0.009$ & -0.054$\pm$0.009 & 0.012$\pm$0.013 & 1.53$\pm$0.20 & -64.23$\pm$3.49\\ 
240.77 & 4.473$\pm$0.009 & $-0.043 \pm 0.009$ & -0.058$\pm$0.009 & 0.013$\pm$0.011 & 1.60$\pm$0.19 & -63.24$\pm$3.63\\ 
240.80 & 4.473$\pm$0.010 & $-0.042 \pm 0.008$ & -0.059$\pm$0.008 & 0.013$\pm$0.013 & 1.62$\pm$0.17 & -62.96$\pm$3.44\\ 
240.83 & 4.472$\pm$0.009 & $-0.043 \pm 0.008$ & -0.057$\pm$0.009 & 0.012$\pm$0.012 & 1.58$\pm$0.20 & -63.47$\pm$3.35\\ 
241.17 & 4.468$\pm$0.008 & $-0.043 \pm 0.008$ & -0.057$\pm$0.008 & 0.012$\pm$0.013 & 1.59$\pm$0.18 & -63.50$\pm$3.35\\ 
241.20 & 4.468$\pm$0.009 & $-0.043 \pm 0.007$ & -0.059$\pm$0.008 & 0.012$\pm$0.014 & 1.64$\pm$0.18 & -62.88$\pm$3.01\\ 
241.23 & 4.467$\pm$0.009 & $-0.043 \pm 0.008$ & -0.058$\pm$0.008 & 0.009$\pm$0.014 & 1.61$\pm$0.18 & -63.19$\pm$3.05\\ 
241.27 & 4.467$\pm$0.009 & $-0.043 \pm 0.007$ & -0.054$\pm$0.008 & 0.008$\pm$0.013 & 1.55$\pm$0.17 & -64.12$\pm$3.09\\ 
241.30 & 4.466$\pm$0.009 & $-0.043 \pm 0.008$ & -0.052$\pm$0.008 & 0.010$\pm$0.013 & 1.51$\pm$0.18 & -64.59$\pm$3.04\\ 
241.33 & 4.466$\pm$0.008 & $-0.043 \pm 0.008$ & -0.054$\pm$0.008 & 0.010$\pm$0.014 & 1.54$\pm$0.19 & -64.17$\pm$3.11\\ 
241.36 & 4.465$\pm$0.008 & $-0.043 \pm 0.009$ & -0.056$\pm$0.008 & 0.010$\pm$0.013 & 1.58$\pm$0.19 & -63.69$\pm$3.66\\ 
241.39 & 4.465$\pm$0.009 & $-0.043 \pm 0.008$ & -0.058$\pm$0.007 & 0.010$\pm$0.013 & 1.61$\pm$0.17 & -63.28$\pm$3.14\\ 
241.42 & 4.464$\pm$0.008 & $-0.043 \pm 0.008$ & -0.060$\pm$0.009 & 0.010$\pm$0.012 & 1.65$\pm$0.20 & -62.68$\pm$3.26\\ 
241.45 & 4.464$\pm$0.008 & $-0.043 \pm 0.008$ & -0.060$\pm$0.008 & 0.011$\pm$0.013 & 1.65$\pm$0.19 & -62.78$\pm$3.17\\ 
241.48 & 4.464$\pm$0.008 & $-0.043 \pm 0.007$ & -0.057$\pm$0.009 & 0.011$\pm$0.013 & 1.59$\pm$0.19 & -63.51$\pm$2.91\\ 
241.52 & 4.463$\pm$0.008 & $-0.043 \pm 0.009$ & -0.054$\pm$0.007 & 0.011$\pm$0.013 & 1.54$\pm$0.18 & -64.26$\pm$3.49\\ 
241.55 & 4.463$\pm$0.007 & $-0.043 \pm 0.008$ & -0.053$\pm$0.008 & 0.011$\pm$0.012 & 1.53$\pm$0.19 & -64.36$\pm$3.45\\ 
241.58 & 4.462$\pm$0.010 & $-0.043 \pm 0.008$ & -0.056$\pm$0.008 & 0.012$\pm$0.012 & 1.58$\pm$0.18 & -63.60$\pm$3.14\\ 
241.61 & 4.462$\pm$0.009 & $-0.043 \pm 0.009$ & -0.057$\pm$0.008 & 0.011$\pm$0.012 & 1.59$\pm$0.20 & -63.52$\pm$3.40\\ 
241.64 & 4.461$\pm$0.010 & $-0.043 \pm 0.008$ & -0.056$\pm$0.008 & 0.011$\pm$0.014 & 1.57$\pm$0.18 & -63.73$\pm$3.41\\ 
241.67 & 4.461$\pm$0.010 & $-0.043 \pm 0.009$ & -0.057$\pm$0.008 & 0.012$\pm$0.014 & 1.60$\pm$0.19 & -63.42$\pm$3.47\\ 
241.70 & 4.461$\pm$0.008 & $-0.043 \pm 0.008$ & -0.056$\pm$0.008 & 0.013$\pm$0.012 & 1.58$\pm$0.18 & -63.58$\pm$3.48\\ 
241.73 & 4.460$\pm$0.008 & $-0.043 \pm 0.009$ & -0.055$\pm$0.009 & 0.013$\pm$0.014 & 1.56$\pm$0.21 & -63.93$\pm$3.48\\ 
241.77 & 4.460$\pm$0.008 & $-0.043 \pm 0.008$ & -0.053$\pm$0.009 & 0.011$\pm$0.014 & 1.53$\pm$0.18 & -64.36$\pm$3.78\\ 
241.80 & 4.459$\pm$0.009 & $-0.043 \pm 0.007$ & -0.052$\pm$0.007 & 0.010$\pm$0.014 & 1.51$\pm$0.17 & -64.71$\pm$2.91\\ 
241.83 & 4.459$\pm$0.009 & $-0.043 \pm 0.007$ & -0.055$\pm$0.008 & 0.010$\pm$0.012 & 1.57$\pm$0.17 & -63.84$\pm$3.09\\ 
241.86 & 4.458$\pm$0.009 & $-0.043 \pm 0.009$ & -0.057$\pm$0.008 & 0.010$\pm$0.011 & 1.60$\pm$0.17 & -63.31$\pm$3.89\\ 
241.89 & 4.458$\pm$0.008 & $-0.043 \pm 0.008$ & -0.055$\pm$0.008 & 0.009$\pm$0.013 & 1.55$\pm$0.16 & -64.02$\pm$3.67\\ 
241.92 & 4.458$\pm$0.009 & $-0.043 \pm 0.008$ & -0.053$\pm$0.007 & 0.008$\pm$0.013 & 1.53$\pm$0.18 & -64.42$\pm$3.13\\ 
241.95 & 4.457$\pm$0.009 & $-0.043 \pm 0.009$ & -0.055$\pm$0.009 & 0.009$\pm$0.013 & 1.55$\pm$0.18 & -64.01$\pm$3.88\\ 
241.98 & 4.457$\pm$0.008 & $-0.043 \pm 0.009$ & -0.054$\pm$0.008 & 0.013$\pm$0.012 & 1.55$\pm$0.18 & -64.03$\pm$3.66\\ 
242.02 & 4.456$\pm$0.008 & $-0.043 \pm 0.009$ & -0.054$\pm$0.007 & 0.014$\pm$0.013 & 1.55$\pm$0.17 & -64.12$\pm$3.62\\ 
242.05 & 4.456$\pm$0.008 & $-0.043 \pm 0.008$ & -0.055$\pm$0.008 & 0.013$\pm$0.012 & 1.56$\pm$0.19 & -63.88$\pm$3.38\\ 
242.08 & 4.455$\pm$0.008 & $-0.043 \pm 0.008$ & -0.055$\pm$0.008 & 0.011$\pm$0.013 & 1.56$\pm$0.19 & -63.86$\pm$3.30\\ 
242.11 & 4.455$\pm$0.008 & $-0.043 \pm 0.009$ & -0.057$\pm$0.008 & 0.011$\pm$0.014 & 1.59$\pm$0.18 & -63.51$\pm$3.60\\ 
242.14 & 4.455$\pm$0.009 & $-0.043 \pm 0.009$ & -0.057$\pm$0.009 & 0.011$\pm$0.013 & 1.60$\pm$0.20 & -63.35$\pm$3.66\\ 
242.17 & 4.454$\pm$0.009 & $-0.043 \pm 0.008$ & -0.055$\pm$0.008 & 0.010$\pm$0.012 & 1.56$\pm$0.16 & -63.92$\pm$3.59\\ 
242.20 & 4.454$\pm$0.008 & $-0.043 \pm 0.009$ & -0.052$\pm$0.008 & 0.011$\pm$0.013 & 1.51$\pm$0.19 & -64.68$\pm$3.53\\ 
242.23 & 4.453$\pm$0.008 & $-0.043 \pm 0.009$ & -0.050$\pm$0.007 & 0.011$\pm$0.012 & 1.47$\pm$0.17 & -65.33$\pm$3.60\\ 
242.27 & 4.453$\pm$0.009 & $-0.043 \pm 0.008$ & -0.050$\pm$0.009 & 0.011$\pm$0.014 & 1.47$\pm$0.18 & -65.32$\pm$3.79\\ 
242.30 & 4.452$\pm$0.008 & $-0.043 \pm 0.009$ & -0.053$\pm$0.008 & 0.011$\pm$0.012 & 1.52$\pm$0.18 & -64.52$\pm$3.93\\ 
242.33 & 4.452$\pm$0.009 & $-0.043 \pm 0.008$ & -0.056$\pm$0.008 & 0.010$\pm$0.013 & 1.58$\pm$0.19 & -63.61$\pm$3.17\\ 
242.36 & 4.452$\pm$0.008 & $-0.043 \pm 0.008$ & -0.056$\pm$0.008 & 0.011$\pm$0.013 & 1.59$\pm$0.20 & -63.54$\pm$3.21\\ 
242.39 & 4.451$\pm$0.008 & $-0.043 \pm 0.008$ & -0.054$\pm$0.007 & 0.013$\pm$0.012 & 1.55$\pm$0.16 & -64.10$\pm$3.42\\ 
242.42 & 4.451$\pm$0.008 & $-0.043 \pm 0.008$ & -0.053$\pm$0.008 & 0.012$\pm$0.014 & 1.53$\pm$0.17 & -64.30$\pm$3.18\\ 
242.45 & 4.450$\pm$0.010 & $-0.043 \pm 0.007$ & -0.054$\pm$0.008 & 0.010$\pm$0.013 & 1.54$\pm$0.17 & -64.24$\pm$2.95\\ 
242.48 & 4.450$\pm$0.008 & $-0.043 \pm 0.008$ & -0.055$\pm$0.009 & 0.011$\pm$0.013 & 1.56$\pm$0.20 & -63.96$\pm$3.53\\ 
242.52 & 4.449$\pm$0.008 & $-0.043 \pm 0.009$ & -0.055$\pm$0.008 & 0.012$\pm$0.014 & 1.56$\pm$0.18 & -63.85$\pm$3.70\\ 
242.55 & 4.449$\pm$0.008 & $-0.043 \pm 0.008$ & -0.055$\pm$0.008 & 0.012$\pm$0.012 & 1.57$\pm$0.19 & -63.80$\pm$2.96\\ 
242.58 & 4.449$\pm$0.009 & $-0.043 \pm 0.008$ & -0.056$\pm$0.008 & 0.013$\pm$0.014 & 1.58$\pm$0.18 & -63.57$\pm$3.11\\ 
242.61 & 4.448$\pm$0.009 & $-0.043 \pm 0.008$ & -0.056$\pm$0.008 & 0.012$\pm$0.012 & 1.59$\pm$0.17 & -63.50$\pm$3.42\\ 
242.64 & 4.448$\pm$0.009 & $-0.043 \pm 0.008$ & -0.058$\pm$0.010 & 0.011$\pm$0.012 & 1.61$\pm$0.20 & -63.25$\pm$3.48\\ 
242.67 & 4.447$\pm$0.008 & $-0.043 \pm 0.008$ & -0.057$\pm$0.008 & 0.012$\pm$0.012 & 1.61$\pm$0.17 & -63.30$\pm$3.34\\ 
242.70 & 4.447$\pm$0.008 & $-0.043 \pm 0.009$ & -0.054$\pm$0.008 & 0.014$\pm$0.011 & 1.55$\pm$0.18 & -64.13$\pm$3.42\\ 
242.73 & 4.446$\pm$0.008 & $-0.043 \pm 0.008$ & -0.051$\pm$0.008 & 0.013$\pm$0.012 & 1.50$\pm$0.19 & -64.88$\pm$3.41\\ 
242.77 & 4.446$\pm$0.008 & $-0.043 \pm 0.008$ & -0.051$\pm$0.009 & 0.011$\pm$0.014 & 1.50$\pm$0.19 & -64.79$\pm$3.85\\ 
242.80 & 4.446$\pm$0.008 & $-0.043 \pm 0.008$ & -0.053$\pm$0.008 & 0.011$\pm$0.014 & 1.53$\pm$0.18 & -64.33$\pm$3.32\\ 
242.83 & 4.445$\pm$0.008 & $-0.043 \pm 0.009$ & -0.055$\pm$0.007 & 0.012$\pm$0.013 & 1.56$\pm$0.17 & -63.93$\pm$3.75\\ 
\hline
\end{longtable}
}

\end{document}